\documentclass[fleqn,usenatbib]{mnras}

\usepackage{newtxtext,newtxmath}
\usepackage[T1]{fontenc}

\DeclareRobustCommand{\VAN}[3]{#2}
\let\VANthebibliography\thebibliography
\def\thebibliography{\DeclareRobustCommand{\VAN}[3]{##3}\VANthebibliography}

\usepackage{graphicx}	% Including figure files
\usepackage{amsmath}	% Advanced maths commands
\usepackage{bm}         % Proper bolding of maths
\usepackage{multirow}   % Allows spreading info across multiple rows in tables
\usepackage{tikz,xcolor,hyperref} % For the orcid icons that Paul implemented
\usepackage{enumitem}  % Allows fine control over indentation and separation in lists 
\usepackage{caption}

\newcommand{\xmm}{{\em XMM}}
\newcommand{\chandra}{{\em Chandra}}
\newcommand{\erosita}{{\em eROSITA}}
\newcommand{\xga}{{\sc Xga}}
\newcommand{\verxga}{v0.4.2}
\newcommand{\xspec}{{\sc xspec}}
\newcommand{\tx}{{$T_{\rm{X}}$}}
\newcommand{\lx}{{$L_{\rm{X}}$}}

\newcommand{\xapa}{{\sc Xapa}}
\newcommand{\clustone}{{SDSSXCS-55}}
\newcommand{\clusttwo}{{SDSSXCS-6955}}

\newcommand{\XXLTxvalues}{97}
\newcommand{\XXLgasmassvalues}{91}

\newcommand{\LOCUSSTxvaluesorginal}{33}
\newcommand{\LOCUSSgasmassvaluesoriginal}{33}
\newcommand{\LOCUSSgasmassvaluesxga}{32}

\newcommand{\LOCUSShydromassvaluesxga}{29}
\newcommand{\LOCUSShydromassvaluesoriginalxgaoverlap}{24}

\newcommand{\totaluniqueobsids}{457}
\newcommand{\totalrejectedobsids}{43}
\newcommand{\instrejectedobsids}{4}

\newcommand{\totaluniqclustersvaltests}{{268}}

\newcommand{\sdssrmmasses}{104}

\definecolor{lime}{HTML}{A6CE39}
\DeclareRobustCommand{\orcidicon}{%
	\begin{tikzpicture}
	\draw[lime, fill=lime] (0,0) 
	circle [radius=0.16] 
	node[white] {{\fontfamily{qag}\selectfont \tiny ID}};
	\draw[white, fill=white] (-0.0625,0.095) 
	circle [radius=0.007];
	\end{tikzpicture}
	\hspace{-2mm}
}

\foreach \x in {A, ..., Z}{%
	\expandafter\xdef\csname orcid\x\endcsname{\noexpand\href{https://orcid.org/\csname orcidauthor\x\endcsname}{\noexpand\orcidicon}}
}

\title[XCS: Automating hydrostatic masses for large samples of clusters I - Methodology]{The \xmm{} Cluster Survey: Automating the estimation of hydrostatic mass for large samples of galaxy clusters I - Methodology, Validation, \& Application to the SDSSRM-XCS sample}

\author[D. J. Turner et al.]
{D. J. Turner$^{1,2}$\thanks{E-mail: turne540@msu.edu (DJT)}\orcidA{},
P. A. Giles$^{2}$\orcidB{},
A. K. Romer$^{2}$\orcidC{}, 
J. Pilling$^{2}$,
T. K. Lingard$^{3,2}$\orcidK{},
R. Wilkinson$^{2}$\orcidD{},
\newauthor
M. Hilton$^{4}$\orcidJ{},
E. W. Upsdell$^{2}$\orcidG{},
R. Al-Serkal$^{2}$,
T. Cheng$^{5,2}$,
R. Eappen$^{2}$,
P. J. Rooney$^{2}$,
S. Bhargava$^{6}$\orcidL{},
\newauthor
C. A. Collins$^{7}$,
J. Mayers$^{2}$,
C. Miller$^{8}$,
R. C. Nichol$^{9}$\orcidM{},
M. Sahl\'en$^{10}$,
P. T. P. Viana$^{11,12}$\orcidE{}
\newauthor
\\
$^{1}$Michigan State University, Physics \& Astronomy Department, East Lansing, Michigan, USA\\
$^{2}$Department of Physics and Astronomy, University of Sussex, Brighton, BN1 9QH, UK \\
$^{3}$Institute of Cosmology and Gravitation, University of Portsmouth\\
$^{4}$University of the Witwatersrand Johannesburg, School of Physics, Private Bag 3, Johannesburg, ZA 2050 \\
$^{5}$Imperial College London, Astrophysics Group, Blackett Laboratory, Prince Consort Road, London, UK SW7 2AZ \\
$^{6}$Observatoire de la C\^ote d'Azur, 96 Bd de l'Observatoire, 06300 Nice, France \\
$^{7}$Astrophysics Research Institute, Liverpool John Moores University, Liverpool Science Park, 146 Brownlow Hill, Liverpool L3 5RF, UK \\
$^{8}$Department of Astronomy, University of Michigan, Ann Arbor, MI 48109 USA \\
$^{9}$School of Maths and Physics, University of Surrey, Guildford, UK \\
$^{10}$Theoretical Astrophysics, Department of Physics and Astronomy, Uppsala University, Box 516, SE- 751 20 Uppsala, Sweden \\
$^{11}$Instituto de Astrof\'isica e Ci\^{e}ncias do Espa\c co, Universidade do Porto, CAUP, Rua das Estrelas, 4150-762 Porto, Portugal \\
$^{12}$Departamento de F\'isica e Astronomia, Faculdade de Ci\^{e}ncias, Universidade do Porto, Rua do Campo Alegre, 687, 4169-007 Porto, Portugal 
}

\date{Accepted XXX. Received YYY; in original form ZZZ}

\pubyear{2024}

\begin{document}
\label{firstpage}
\pagerange{\pageref{firstpage}--\pageref{lastpage}}
\maketitle

% Abstract of the paper
\begin{abstract}
We describe features of the X-ray: Generate and Analyse (\xga{}) open-source software package that have been developed to facilitate automated hydrostatic mass ($M_{\rm hydro}$) measurements from {\em XMM} X-ray observations of clusters of galaxies. This includes describing how \xga{} measures global, and radial, X-ray properties of galaxy clusters.
We then demonstrate the reliability of \xga{} by comparing simple X-ray properties, namely the X-ray temperature and gas mass, with published values presented by the \xmm{} Cluster Survey (XCS), the Ultimate {\bf X}MM e{\bf X}traga{\bf L}actic survey project (XXL), and the Local Cluster Substructure Survey (LoCuSS). \xga{} measured values for temperature are, on average, within 1\% of the values reported in the literature for each sample. \xga{} gas masses for XXL clusters are shown to be ${\sim}$10\% lower than previous measurements (though the difference is only significant at the $\sim$1.8$\sigma$ level), LoCuSS $R_{2500}$ and $R_{500}$ gas mass re-measurements are 3\% and 7\% lower respectively (representing 1.5$\sigma$ and 3.5$\sigma$ differences). Like-for-like comparisons of hydrostatic mass are made to LoCuSS results, which show that our measurements are $10{\pm}3\%$ ($19{\pm}7\%$) higher for $R_{2500}$ ($R_{500}$). The comparison between $R_{500}$ masses shows significant scatter.  Finally, we present new $M_{\rm hydro}$ measurements for \sdssrmmasses{} clusters from the SDSS DR8 redMaPPer XCS sample (SDSSRM-XCS). Our SDSSRM-XCS hydrostatic mass measurements are in good agreement with multiple literature estimates, and represent one of the largest samples of consistently measured hydrostatic masses.  We have demonstrated that \xga{} is a powerful tool for X-ray analysis of clusters; it will render complex-to-measure X-ray properties accessible to non-specialists.
\end{abstract}

% Select between one and six entries from the list of approved keywords.
% Don't make up new ones.
\begin{keywords}
X-rays: galaxies: clusters -- galaxies: clusters: intracluster medium -- galaxies: clusters: general -- methods: data analysis -- methods: observational
\end{keywords}

%%%%%%%%%%%%%%%%%%%%%%%%%%%%%%%%%%%%%%%%%%%%%%%%%%

%%%%%%%%%%%%%%%%% BODY OF PAPER %%%%%%%%%%%%%%%%%%
\section{Introduction}
\label{sec:intro}

Galaxy clusters are the most massive virialized structures in the Universe. They formed through the collapse of the primordial density field, and as such are a useful way to investigate the evolution of the Universe through the measurement of cosmological parameters. The mass of a galaxy cluster is split into three main components \cite{massbudget}; the dark matter halo (87\%), the intra-cluster medium (7\%), and the component galaxies (3\%); where the intra-cluster medium (ICM) is a high-temperature, low-density plasma largely made up of ionised hydrogen. Just as the formation of clusters makes them useful for investigating cosmology, the nature of the ICM makes them ideal astrophysical laboratories.

Cosmological parameters can be derived using galaxy clusters via a variety of methods.  For example, cosmological parameters have been constrained using samples of X-ray selected clusters by measuring the mass function of clusters \citep[e.g.][]{vikhlinincosmo,schellenbergercosmo}.  Therefore, one of the key properties that must be measured is the cluster mass \citep[see][for a recent review]{pratt_review}.  The Dark Energy Survey (DES) used galaxy clusters detected in the first year of DES observations (DESY1) to constrain cosmological parameters \citep{desclustercosmo} using a weak-lensing mass calibration \citep{desy1_wl_mor} and the number density of clusters.  The weak lensing mass calibration for the DESY1 analysis took the form of a mass-observable relation (MOR), where the observable was the cluster richness.  The richness here, symbolised by $\lambda$, is a probabilistic measure of the number of galaxies in the cluster, estimated from the red-sequence Matched-filter Probabilistic Percolation cluster finder \citep[or redMaPPer,][]{redmapperdessv}.  However, a drawback of the DESY1 analysis is the use of stacked weak lensing masses.  Due to the use of a stacked analysis, information on the intrinsic scatter is lost.  To enable the next generation of cluster cosmology, large scale optical/near-infrared galaxy cluster surveys will need the scatter, and normalisation, of mass-richness relations to be well calibrated. This includes the cosmology analysis that will be performed using clusters detected from the upcoming Vera Rubin Observatory’s Legacy Survey of Space and Time \citep[LSST; e.g. see Figure~G2 in][]{2018arXiv180901669T}. 

One potential method to infer the scatter of these MORs, is through the hydrostatic equilibrium mass ($M_{\rm{hydro}}$) method, using X-ray data to infer the total mass of the cluster from the temperature and gas density profiles.  Of particular use to derive X-ray $M_{\rm{hydro}}$ values is the {\em XMM-Newton} telescope (hereafter {\em XMM}), whose field of view (FoV), high effective area, and large public archive of observations makes it the ideal for these measurements. Previous $M_{\rm{hydro}}$ measurements using \xmm{} include, but are not limited to, \cite{2010A&A...524A..68E}; \cite{xmm_group_masses}; \cite{clash_masses}; \cite{high-z_hydro}; \cite{xcop_hydro}; \cite{lovisarimasses}; \cite{2023MNRAS.520.6001P}. The largest of these studies, \cite{lovisarimasses}, yielded 120 $M_{\rm{hydro}}$ values.
For completeness, we note that hydrostatic masses have also been measured using other (than \xmm{}) X-ray instruments, e.g. \citet[][]{ascatempstruct,ascaabell2029,ascahydromass,vikhlininmass,sungroupshydro,clash_masses,gileshydro,hicosmo_masses,logan22,pointysanders}.  The largest of these studies, \cite{hicosmo_masses}, yielded 64 $M_{\rm{hydro}}$ values. Furthermore, recent efforts have been made to develop methods of estimating X-ray based masses from clusters detected in the {\em eROSITA} All Sky Survey \citep{erass_masses}.

Gas density profiles are an important part of measuring the hydrostatic mass, but they provide a great deal of information in their own right; they are also easier to measure than hydrostatic masses. They have been measured using various X-ray telescopes, including \xmm{} and \chandra{} \citep[][]{crostondens,acceptprofs,chandraxmmdensprofs}. The largest of these studies \citep[][]{acceptprofs}, measured profiles for 239 galaxy clusters. Other X-ray observatories have also been used, including Suzaku \citep[][]{suzakudensprof}, and ROSAT \citep[][]{rosatdens,rosatstackdens}.

The outline of the paper is as follows. Section \ref{sec:methodology} introduces the features of the X-ray: Generate and Analyse (\xga{}) software package that have been developed to facilitate automated hydrostatic masses ($M_{\rm hydro}$). In Section \ref{sec:validation}, we demonstrate the reliability of the \xga{} measurements by comparing with published values presented by the \xmm{} Cluster Survey \citep[][XCS hereafter]{xcsfoundation}, the Local Cluster Substructure Survey\footnote{\href{http://www.sr.bham.ac.uk/locuss/}{LoCuSS Website - http://www.sr.bham.ac.uk/locuss/}} (hereafter LoCuSS), and the Ultimate {\bf X}MM e{\bf X}traga{\bf L}actic (hereafter XXL) survey project \citep{xxl1}. In Section \ref{sec:results}, we present new $M_{\rm hydro}$ measurements of clusters in SDSS DR8 redMaPPer XCS sample. Finally, in Section \ref{sec:conclusions}, we present our conclusions and a discussion of the next steps of this work. 

The analysis code, samples, and outputs are available in a GitHub repository\footnote{\label{foot:gitrepo}\href{https://github.com/DavidT3/XCS-Mass-Paper-I-Analysis}{Code/Samples - https://github.com/DavidT3/XCS-Mass-Paper-I-Analysis}}. In Section \ref{sec:validation}, we adopt the cosmology parameters used in  each of the original analyses to which we compare, i.e. $\Omega_{\rm{M}}$=0.3, $\Omega_{\Lambda}$=0.7, and $H_{0}$=70 km s$^{-1}$ Mpc$^{-1}$ for the XCS and LoCuSS samples (Sections~\ref{subsubsec:SDSSRMsim} and \ref{subsubsec:LoCuSSsim} respectively), and $\Omega_{\rm{M}}$=0.282, $\Omega_{\Lambda}$=0.719, and $H_{0}$=69.7 km s$^{-1}$ Mpc$^{-1}$ for the XXL sample (Section~\ref{subsubsec:XXLsim}) \citep[i.e. the WMAP9 values in ][]{wmap9cosmo}.  In Section \ref{sec:results}, we again use $\Omega_{\rm{M}}$=0.3, $\Omega_{\Lambda}$=0.7, and $\rm{H}_{0}$=70 km s$^{-1}$ Mpc$^{-1}$.

\section{Methodology}
\label{sec:methodology}

The overarching aim of this work is to provide an independent mass calibration for optically selected cluster samples for the purpose of cosmological parameter estimation. The masses are estimated using X-ray observations from \xmm{} under the assumptions of spherical symmetry and hydrostatic equilibrium \citep[see e.g.,][for a derivation]{1980ApJ...241..552F}, following the equation

\begin{equation}
    M_{\rm hydro}(<r) = - \dfrac{k_{B}r^2}{\rho_{g}(r)\mu m_{u}G} \bigg[\rho_{g}(r)\dfrac{dT(r)}{dr} + T(r)\dfrac{d\rho_{g}(r)}{dr}\bigg],
    \label{eq:hydromass}
\end{equation}
where $r$ is the radius within which the mass is being measured, $\rho_{g}(r)$ is the intracluster gas density profile, $T(r)$ is the gas temperature profile, $\mu=0.61$ is the mean molecular weight, $m_{u}$ is the atomic mass unit, $G$ is the gravitational constant, and $k_{B}$ is the Boltzmann constant. The quantities that need to be estimated from the \xmm{} data are $T(r)$ and $\rho_{g}(r)$. It is important to note that these are the 3 dimensional quantities, rather than the projected (i.e. what is observed) values. 

Many steps are needed to go from a raw \xmm{} observation to an estimate of $M_{\rm hydro}$.  Therefore, a secondary goal of this work is to streamline those steps into a single, self-contained workflow. Doing so ensures the consistency of the data products and allows for computational speed-ups (e.g. by making use of multi-threading on multi-core machines).
The measurement tools for $M_{\rm hydro}$ that we use herein are part of the X-ray: Generate and Analyse (\xga\footnote{\href{https://github.com/DavidT3/XGA}{\xga{} GitHub - https://github.com/DavidT3/XGA}}) software suite. \xga{} is a generalised X-ray analysis tool, capable of investigating any X-ray source that has been observed by \xmm{}. It was introduced in \cite{xgapaper} and is being used in a growing number of scientific applications \citep[e.g. ][]{denishapaper,efedsxcs,agnxga}. It is also listed on the Astrophysics Source Code Library \citep[ASCL;][]{asclxga}. This work makes use of \xga{} \verxga{} for all analyses, and to aid readability, the technical details about which parts of \xga{} are used for the different analyses in this work are described in Appendix~\ref{app:xgacommclass}.

In this section, we describe each step involved in \xga{} $M_{\rm hydro}$ measurements.
Section~\ref{subsec:xcsdata} explains the data inputs required to initiate \xga{}.
Image generation and masking of contaminating sources are described in Sections~\ref{subsec:xgaphot} and~\ref{subsec:xgacontam} respectively.  Manual checks of the data are outlined in Section~\ref{subsec:manualdetreg}.  Section~\ref{subsec:xgaspecs} details the generation and fitting of X-ray spectra.  Correcting for the {\em XMM} point-spread function (PSF) is detailed in Section~\ref{subsec:psfcorr}.  The generation of emissivity profiles, density profiles, temperature profiles and mass profiles are outlined in Sections~\ref{subsec:SBandEm},~\ref{subsec:densityProf},~\ref{subsec:gastempprof} and~\ref{subsec:hydromassprof} respectively. To illustrate the steps, we use two example clusters: \clustone{} and \clusttwo{} ($z=0.119$ and $z=0.223$ respectively). Both clusters were part of the \cite{xcsgiles} study (see Section~\ref{subsubsec:SDSSRMsim}), but have significantly different signal to noise ratios and off axis locations in their respective \xmm{} observations. Their properties are summarised in Table~\ref{tab:exampclusters}.

\begin{table*}
\begin{center}
\caption[]{{Properties of the two galaxy clusters selected as demonstration cases for \xga{}. These were selected from the SDSSRM-XCS sample presented by \cite{xcsgiles}. The coordinates are for the XCS \xapa{} defined value for the UDC$^{\ref{foot:udc}}$. This, and the $R_{2500}$ and $R_{500}$ values, in units of kpc, were taken from \cite{xcsgiles}. The redshift and richness values were taken from \cite{redmappersdss} (where the numeric value in the name is the identifier in the \cite{redmappersdss} RM catalogue). The XCS3P measured $T_{\rm X}$ values are taken from \cite{xcsgiles}, and are given in $R_{2500}$ and $R_{500}$ apertures}\label{tab:exampclusters}}
\vspace{1mm}
\begin{tabular}{ccccccccc}
\hline
\hline
Name & RA & Dec & $z_{\rm{RM}}$ & $\lambda_{\rm{RM}}$ & $R_{\rm{2500}}$ &$R_{\rm{500}}$ & $T^{\rm{XCS3P}}_{\rm{X,2500}}$ &$T^{\rm{XCS3P}}_{\rm{X,500}}$  \\
 & (deg) & (deg) & & & (kpc) & (kpc) & (keV) & (keV) \\
\hline
\hline 
\clustone{} & 227.550 & 33.516 & 0.119 
& $99.80{\pm}3.74$\
& $562.3^{+7.5}_{-7.5}$ & $1237.6^{+20.1}_{-20.1}$ & $7.00^{+0.10}_{-0.10}$ & $6.70^{+0.08}_{-0.08}$ 
\\
\hline
\clusttwo{} & 36.455 & -5.894 & 0.223 
& $32.12{\pm}2.83$
& $321.9^{+24.4}_{-18.1}$ & $669.0^{+62.0}_{-47.9}$ & $2.89^{+0.38}_{-0.28}$ & $2.41^{+0.39}_{-0.29}$ \\
\hline
\end{tabular}
\end{center}
\end{table*}

\begin{figure*}
    \centering
    \includegraphics[width=1.0\textwidth]{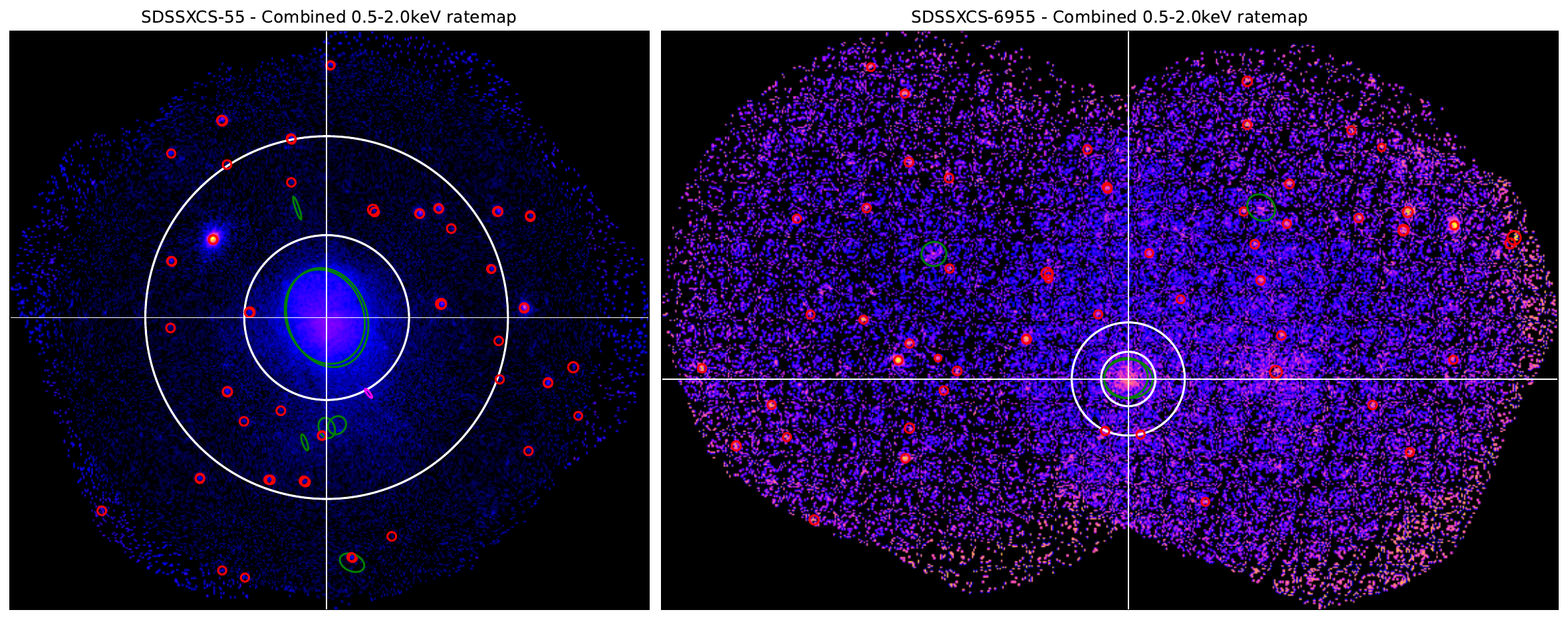}
    \caption[XAPA demo]{\xmm{} 0.5-2.0~keV stacked ratemaps of \clustone{} (left) and \clusttwo{} (right), which have been selected to demonstrate our methodology. Point sources are highlighted by red circles and extended sources by green ellipses.  The white cross-hair indicates the cluster centroid, and the solid inner and outer while circles indicate the $R_{2500}$ and $R_{500}$ radii respectively (see Table~\ref{tab:exampclusters} for input values).   
    }
    \label{fig:xapaexample}
\end{figure*}

\subsection{Initiating \xga{}}
\label{subsec:xcsdata}

Throughout this work, we use public {\em XMM} data from the {\em XMM} Science Archive\footnote{\href{http://nxsa.esac.esa.int/nxsa-web/}{XSA - http://nxsa.esac.esa.int/nxsa-web/}} that has been pre-processed to produce cleaned event files and region lists. The region lists encode information about the source centroid, size, and shape. All pre-processing used herein relies on the the XCS methodology that is fully described in \cite{xcsmethod} and \cite{xcsgiles}. In brief, the initial EPIC (MOS and PN) data were processed with v14.0.0 of the \xmm{} Science Analysis Software \citep[SAS\footnote{\href{https://www.cosmos.esa.int/web/xmm-newton/sas}{SAS - https://www.cosmos.esa.int/web/xmm-newton/sas}};][]{sas}, using {\tt EMCHAIN} and {\tt EPCHAIN} functions to generate event lists. Following this, the event lists were screened for periods of high background and then individual (PN, MOS1, MOS2) and merged (PN+MOS1+MOS2) EPIC images (and corresponding exposure maps) were then generated. All images/maps have a pixel size of 4.35\arcsec. 
Next, X-ray source detection is performed using a custom version of {\sc wavdetect} \citep[][]{wavdetect} called \xapa{} (for XCS Automated Pipeline Algorithm). Once sources in an image have been located, \xapa{} classifies them as either point-like or extended sources. Region lists of sources are created for each \xmm{} observation (a given source can appear in multiple region lists if there are overlapping observations). Once all the observations have been processed, a master source list (MSL), with duplicates removed, is generated. 
To date, 12,582 observations have been processed by XCS, covering 1,068 non-overlapping square degrees of sky, and yielding 400,225 X-ray source detections in the 0.5-2.0~keV band.

% This briefly sets the footnotes to 'symbol mode', so that we can use the dagger for a footnote
\renewcommand{\thefootnote}{\fnsymbol{footnote}}

In addition to the XCS supplied cleaned event files and region lists, the following information is required to initiate the \xga{} analysis of a cluster: a coordinate for the cluster centre, the redshift, at least one radius to define the analysis region, and a set of values for the cosmological parameters $\Omega_M$, $\Omega_\Lambda$, and $H_0$. With regard to the cluster centre, this depends on the selection method (e.g. it could be defined by the brightest galaxy, rather than a feature in an X-ray image) and on the analysis method (e.g. it could be defined by the peak of the X-ray surface brightness or by the weighted centroid). Given this ambiguity, we henceforth refer to the user-defined coordinate (or UDC\footnote[2]{\label{foot:udc}The User Defined Centroid (UDC) is a coordinate input by the user, rather than a peak or centroid measured by \xga{} itself.}), rather than to the ``cluster centre''. 

% This puts the footnote back to normal numbering
\renewcommand{\thefootnote}{\arabic{footnote}}

With regard to the analysis region radius, this can be an overdensity radius\footnote{Radius at which the average density of the enclosed cluster is equal to $\Delta\rho_{c}(z)$, where $\Delta$ is the overdensity factor and $\rho_{c}(z)$ is the critical density of the Universe at the cluster redshift.}, e.g. $R_{500}$, or a proper radius with no associated physical significance (e.g. 300~kpc).
%a cosmological model from Astropy \citep[][]{astropy1,astropy2}.

 For our example clusters, \clustone{} and \clusttwo{} (and for the analysis presented in Section~\ref{sec:results}), the UDC$^{\ref{foot:udc}}$, the redshift, and the analysis region radii ($R_{2500}$, $R_{500}$) were taken from the data tables in \cite{xcsgiles}. \cite{xcsgiles} analysed these clusters using the XCS Post Processing Pipeline (XCS3P).  As per \cite{xcsgiles}, the cosmological parameters were set to $\Omega_{\rm{M}}$=0.3, $\Omega_{\Lambda}$=0.7, and $H_{0}$=70 km s$^{-1}$ Mpc$^{-1}$.

To ensure all relevant data sets are used during the analysis of a given cluster, \xga{} explores the full set of XCS processed data 
to retrieve any \xmm{} observations meeting these criteria; an aimpoint within 30\arcmin{} of the input UDC$^{\ref{foot:udc}}$, and >70\% of the chosen analysis region falling on active regions of the detectors. Two observations each were retrieved for our example clusters (Table~\ref{tab:exampclusters}), \clustone{} (0149880101 \& 0303930101) and \clusttwo{} (0404965201 \& 0677600131).

\subsection{Generating image and exposure maps}
\label{subsec:xgaphot}

Once the relevant {\em XMM} data have been retrieved, see Section~\ref{subsec:xcsdata}, \xga{} generates images, exposure maps, and ratemaps\footnote{The image divided by the exposure map.}.  For this, \xga{} interfaces with the \texttt{SAS} \texttt{evselect} and \texttt{eexpmap} tools. These images and maps are essential to the derivation of $M_{\rm{hydro}}$ as they are used to generate of surface brightness profiles, which in turn are used to estimate $\rho_g(r)$ (see Section~\ref{subsec:SBandEm}).  Images are also useful for inspection throughout the analysis process e.g., to check masked regions (see Section~\ref{subsec:manualdetreg}). 

We note that, in principle, the images already existing in the XCS archive (those on which \xapa{} was run) could have been used by \xga{} for the analyses herein. However, in practice, generation of fresh images is worthwhile because it simplifies the process of merging multiple observations, and gives additional freedom in the current analysis (e.g., selecting an energy range; XCS only stores images in the 0.5-2.0~keV and 2.0-10.0~keV ranges).

In Figure~\ref{fig:xapaexample}, we show the \xga{} generated \xmm{} images of our example clusters, \clustone{} and \clusttwo{}; these are the stacked images of every EPIC instrument for every usable observation of the clusters. The \xapa{} (Section~\ref{subsec:xcsdata}) defined source regions are overlaid; with red (point source), green (extended source), or magenta (extended source similar in size to the PSF) outlines. The UDC$^{\ref{foot:udc}}$ is shown with a white cross. The analysis regions ($R_{2500}$ and $R_{500}$) are shown with white solid  outlines. 
Note that, in these examples, two observations have been stacked into a composite image. As each observation has its own source region list, multiple source outlines can sometimes be seen in the overlap regions.

\subsection{Automated source masking}
\label{subsec:xgacontam}

\begin{figure*}
    \centering
    \includegraphics[width=1.0\textwidth]{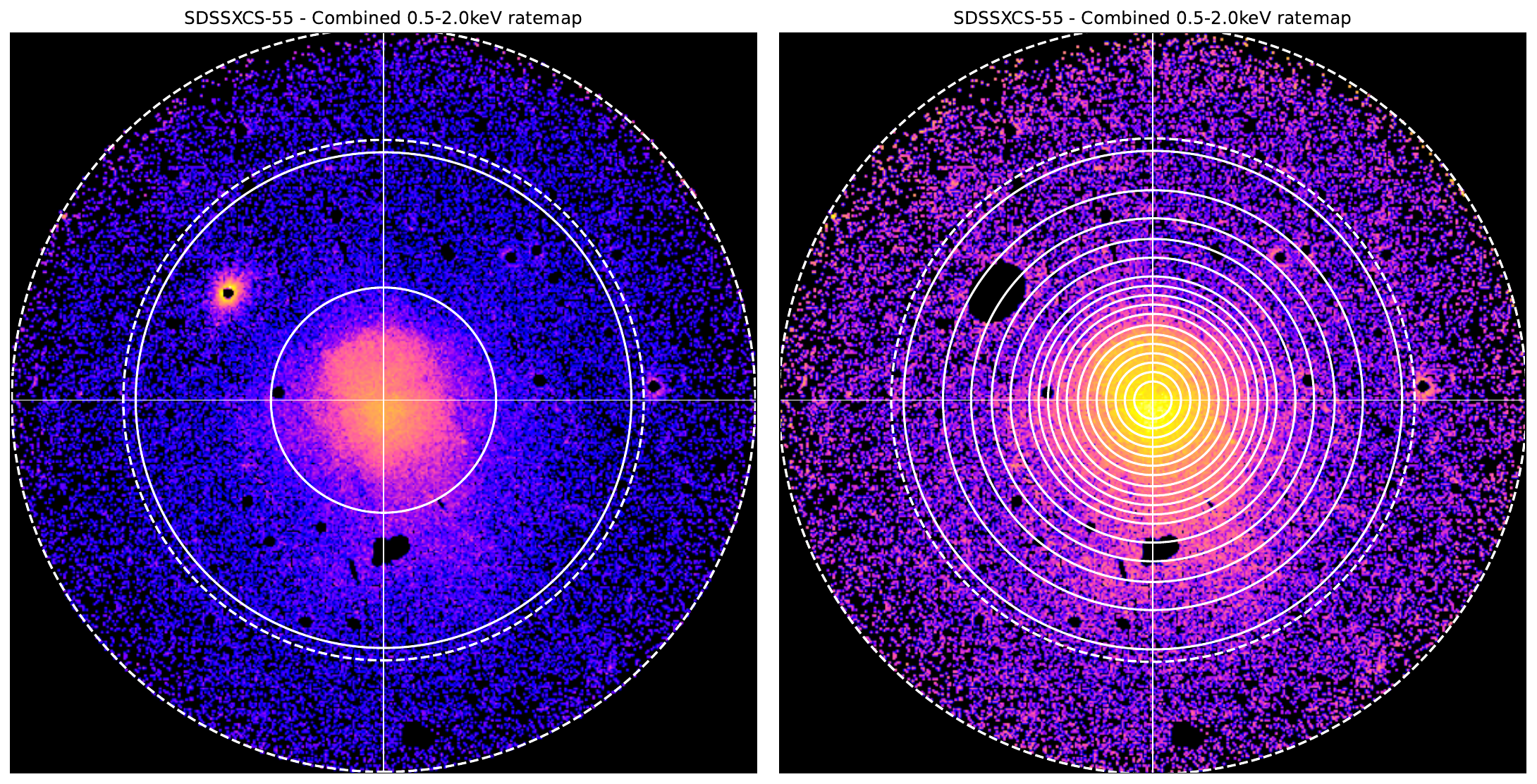}
    \caption[image auto mask]{An \xmm{} 0.5-2.0~keV stacked ratemap of  \clustone{}. The white cross-hairs indicate the UDC$^{\ref{foot:udc}}$. 
    Left: 
    The solid inner and outer circles indicate the location of $R_{2500}$ and $R_{500}$ respectively. The dashed annulus indicate 1.05-1.5$R_{500}$ (the background region used in the global spectral analysis, see Section~\ref{subsec:xgaspecs}).  An automated mask, to remove contaminating sources identified by {\sc xapa} has been applied (black areas, see Section~\ref{subsec:xgacontam}). Right: The default mask has been adjusted by hand to remove more emission from the bright point source in the top left (see Section~\ref{subsec:manualdetreg}). The solid (dashed) circles indicate the source (background) annuli used during the radial spectral fitting used to measure projected temperature profiles (see Sections~\ref{subsec:xgaspecs} and \ref{subsec:gastempprof}).}
    \label{fig:immask}
\end{figure*}

As described above (Section~\ref{subsec:xcsdata}), every XCS processed image has an associated list of detected source regions. Most of these will not be associated with the cluster of interest and need to be masked from both the image files (Section~\ref{subsec:xgaphot}), used during surface brightness fitting (Section~\ref{subsec:SBandEm}), and from the cleaned event lists (Section~\ref{subsec:xcsdata}), used during spectral analysis (Section~\ref{subsec:xgaspecs}). In almost all cases, the only source region not masked by default is the one associated with the cluster of interest (see below for two exceptions). This source is identified as the one containing the location of the UDC$^{\ref{foot:udc}}$. For the analysis herein, we apply an additional filter: the source must also have been classed as extended by the \xapa{} pipeline.  The black areas in Figure~\ref{fig:immask} (left)  show the region automatically masked for \clustone{}. This figure is a zoom into the $3R_{500} \times 3R_{500}$ region of Figure~\ref{fig:xapaexample} (left).

Below are two exceptions to the masking process detailed above:
\begin{itemize}[leftmargin=*, labelindent=2pt, itemsep=1ex]
 \item[--] The treatment of \xapa{} point-like sources detected within 0.15$R_{\rm{analysis}}$ of the cluster centroid. These sources are not masked because \xapa{} occasionally misidentifies cool cores of clusters as point sources.
 \item[--] The treatment of overlapping \xapa{} extended source regions. If the cluster of interest was observed in two or more XMM pointings, then the associated \xapa{} extended sources may not fully overlap. This is more likely if the observations are offset from each other and/or that have differing exposure times. Therefore, if the respective source centres are within a projected distance of $<R_{500}$ of the input cluster centroid, then none of those source regions are masked. 
\end{itemize}

\subsection{Manual data checks}
\label{subsec:manualdetreg}

Once the images have been generated (Section~\ref{subsec:xgaphot}) and the default masks applied (Section~\ref{subsec:xgacontam}), the next step involves manual intervention via eye-ball checks to {\em i}) identify observations that were not suitable for further analysis, {\em ii}) increase the size of masked regions if the default mask was not large enough, and {\em iii}) add masked regions for sources missed by \xapa. With regard to {\em i}), this process is similar to that described in Section~2.3 of \cite{xcsgiles}. It is used to remove from further analysis observations with abnormally high background levels \citep[e.g. Figure~A1(a) of][]{xcsgiles}, and those corrupted by a very bright point source \citep[such sources produce artefacts in the {\em XMM} images including readout streaks and ghost images of the telescope support structure; e.g. Figure~A1(b) of][]{xcsgiles}. Of the \totaluniqueobsids{} \xmm{} observations identified as being relevant to the clusters in this paper, \totalrejectedobsids{} were rejected entirely, and \instrejectedobsids{} had data from one or more (of the 3 EPIC) instruments rejected.

With regard to {\em ii}), occasionally the \xapa{} defined region is not large enough to encapsulate all the emission from a bright point source, or from a neighbouring (but physically distinct) extended source. This can impact subsequent analysis if the source falls in either within $R_{500}$ of the cluster UDC$^{\ref{foot:udc}}$ or within the background annulus. For the \cite{xcsgiles} analysis, this extra masking step was cumbersome and time-consuming. So, in \xga{}, a GUI was developed that makes it simple to interact with, and modify existing regions. With regard to {\em iii}), there is a rare \xapa{} failure mode whereby it fails to detect some of the sources in a given observation. Using the GUI, it is easy to add new regions if a source or artefact (visible by eye in the image) has been missed by \xapa{}. 

Combined images (all observations and instruments that passed our flare checks) for each of the galaxy clusters in this work were examined, and adjustments to regions were made. We altered or added regions for 60 (40\%) of the SDSSRM-XCS sample, 44 (44\%) of the XXL-100-GC sample, and 26 (56\%) of the LoCuSS High-\lx{} sample. Figure~\ref{fig:immask}  (right) shows part of the \xmm{} observation containing \clustone{}. This Figure highlights where a source mask has been expanded from its default (\xapa) size. 

\subsection{Generating and fitting X-ray spectra}
\label{subsec:xgaspecs}

Spectral analysis is essential to the estimation of hydrostatic masses, both for the derivation of $\rho_g(r)$ and $T(r)$. In the case of $\rho_g(r)$, only a global (cluster wide) spectral analysis is needed, whereas for $T(r)$, spectral analysis in radial bins is required. 

For the global spectral analysis of our example clusters, the analysis region (radius and UDC$^{\ref{foot:udc}}$) is defined by the user as an initiation input to \xga{} (Section~\ref{subsec:xcsdata}). A background region also needs to be defined. In the implementation of \xga{} used herein, the background region is an annulus. For our example clusters, \clustone{} and \clusttwo{} (and for the analysis presented in Section~\ref{sec:results}), the inner and outer radii of the annulus are set at 1.05R$_{500}$ and 1.5R$_{500}$ respectively; for $R_{2500}$ measurements they are set at 2$R_{2500}$ and 3$R_{2500}$, and for 300~kpc measurements (for the XXL-100-GC sample, Section~\ref{subsubsec:XXLsim}) they are set to 3.33$R_{\rm{300~kpc}}$ and 5$R_{\rm{300~kpc}}$.
 
For the radial spectral analysis, \xga{} uses a series of $N$ circular source apertures with increasing radii, out to a user defined outer radius, $R_{\rm{outer}}$ and centered on the UDC$^{\ref{foot:udc}}$. The innermost aperture is a full circle, the others are annuli. The user defines a minimum annulus width (also the inner circle radius) in arcseconds, $\Delta\theta_{\rm min}$. This minimum is set to account for the PSF of the instrument. For all the cluster analyses herein, we set this to be $\Delta\theta_{\rm min}$=20\arcsec{} (roughly twice the FWHM of the \xmm{} EPIC-PN on-axis PSF, see Section~\ref{subsec:psfcorr}). 
If the specified $R_{\rm{outer}}$ value for a given cluster, when converted to units of arcseconds ($\theta_{\rm{outer}}$), is not an integer multiple of $\Delta\theta_{\rm min}$, it is adjusted (outwards) accordingly, to become $\theta_{\rm{outer}}'$. A background region is then defined and used for all the bins. For the radial analysis of our example clusters, \clustone{} and \clusttwo{} (and the analysis presented in Section~\ref{sec:results}), it was set to be an annulus of 1.05-1.5~$\theta_{\rm{outer}}'$. 

The widths of the bins used in the spectral analysis are set through an iterative process. This starts with determining whether $\theta_{\rm{outer}}'$ is at least $4\times\Delta\theta_{\rm min}$. If not, then the process stops (because it is not realistic to generate a $T(r)$ profile from less than 4 bins). If it is exactly $4\times\Delta\theta_{\rm min}$, then 4 bins (each of $\theta_{\rm min}$ in width) are used in the analysis. If it is $5\times\Delta\theta_{\rm min}$, then some bins may be expanded (to $2\times\Delta\theta_{\rm min}$, $3\times\Delta\theta_{\rm min}$ etc.) to improve the signal to noise. The width of a given bin expands (inwards) until either a) the user defined minimum number of background subtracted counts is reached, or b) the number of bins drops to 4.
The number of bins defined in this way will depend on the quality of the detection and on the $R_{\rm outer}$ (and hence projected size) of the cluster.

The way that the annular spectra are radially binned can have a significant effect on the temperature profiles that are created from them, and different criteria may be used to decide on the binning \citep[see e.g.,][]{2023arXiv231110397C}. The goal is to make the extraction region large enough that sufficient X-ray counts are present to constrain spectral properties (e.g. temperature), whilst maintaining a good spatial resolution.

Achieving a minimum number of counts per spectral annulus is not always sufficient to guarantee a `good' temperature profile. In an effort to mitigate potential issues, we generate two sets of temperature profiles, with different targeted minimum counts per bin (1500 and 3000 counts). As 1500 counts should be sufficient for a well constrained $T_{\rm X}$ value \citep[see Figure~16 in][]{xcsmethod}, preference is given to profiles measured from those spectra. However, if there is a problem with the 1500 count-binned temperature profile, we instead use the 3000 count-binned profile. Such problems include:
\begin{itemize}[leftmargin=*, labelindent=2pt, itemsep=1ex]
    \item The spectral fitting process failing to converge for some annuli. Even if only one annulus failed in this manner, the entire temperature profile is unusable.
    \item Annular temperature values have very poorly constrained uncertainties, this can make the deprojection process quite unstable, and cause problems when fitting a temperature profile model to the final 3D temperature profile.
    \item Model fits to the deprojected, three-dimensional, temperature profile resulting in unphysical mass profiles, where the hydrostatic mass radial profile is not monotonically increasing. 
\end{itemize}

For the clusters analysed herein, the number of annular bins varied from $N=4$ (the defined minimum) to $N=22$. For our example clusters, it was $N=17$ and $N=4$ for \clustone{} and \clusttwo{} respectively.

Once the source and background regions have been defined, spectra are generated for every camera-plus-observation combination (and in each bin for the radial analysis) of the respective galaxy cluster.  

For our example clusters, \clustone{} and \clusttwo, this corresponds to 6 (132) and 6 (24) spectra respectively for the global (radial) analyses; the annular spectra for this example were generated with a minimum of 1500 counts. \xga{} uses the \texttt{SAS} \texttt{evselect} tool to create the initial spectra, selecting events with a FLAG value of 0, and a PATTERN of ${\leq}4$ for PN and ${\leq}12$ for MOS data. The SAS \texttt{specgroup} tool is used to re-bin each spectrum so that there is minimum number of
counts per channel (in the analysis herein, we set that to be 5, though it can be configured by the user). 

The spectral analysis requires response curves (known as ancillary response files; ARFs) to proceed. These are calculated for each spectrum individually. For this, we use a detector map generated with {\tt x} and {\tt y} bin sizes set to 200 detector coordinate pixels, and with the same event selection criteria as the spectrum. We preferentially select an image from another instrument, if available, e.g. a PN image for a MOS detector map, and vice versa. This helps to mitigate the effects of chip gaps.

\xga{} uses \xspec{} \citep[][]{xspec} to fit emission models to the spectra. For the analyses herein\footnote{\xga{} can use a range of other emission models if the user prefers.}, we use an absorbed plasma emission model \citep[\texttt{constant${\times}$tbabs${\times}$apec;}][]{tbabs, apec}. The multiplicative factor represented by \texttt{constant} helps to account for the difference in normalisation between spectra being fitted simultaneously. The $\rm{n_H}$ values required by \texttt{tbabs} are retrieved using the HEASoft \texttt{nh} command\footnote{\cite{nh} data are used by the \texttt{nh} tool.}. The starting metal abundance value ($Z$) is defined by the user. For the analysis herein, $Z=0.3Z_\odot$ was used.
The redshift is always fixed at the input value during \xga{} fitting. The $\rm{n_H}$ and $Z$ value can be fixed or left free depending on the use case. For our example clusters, \clustone{} and \clusttwo{}, and the analysis in Section~\ref{sec:results}, they were fixed.

\begin{figure}
    \centering
    \includegraphics[width=1.\columnwidth]{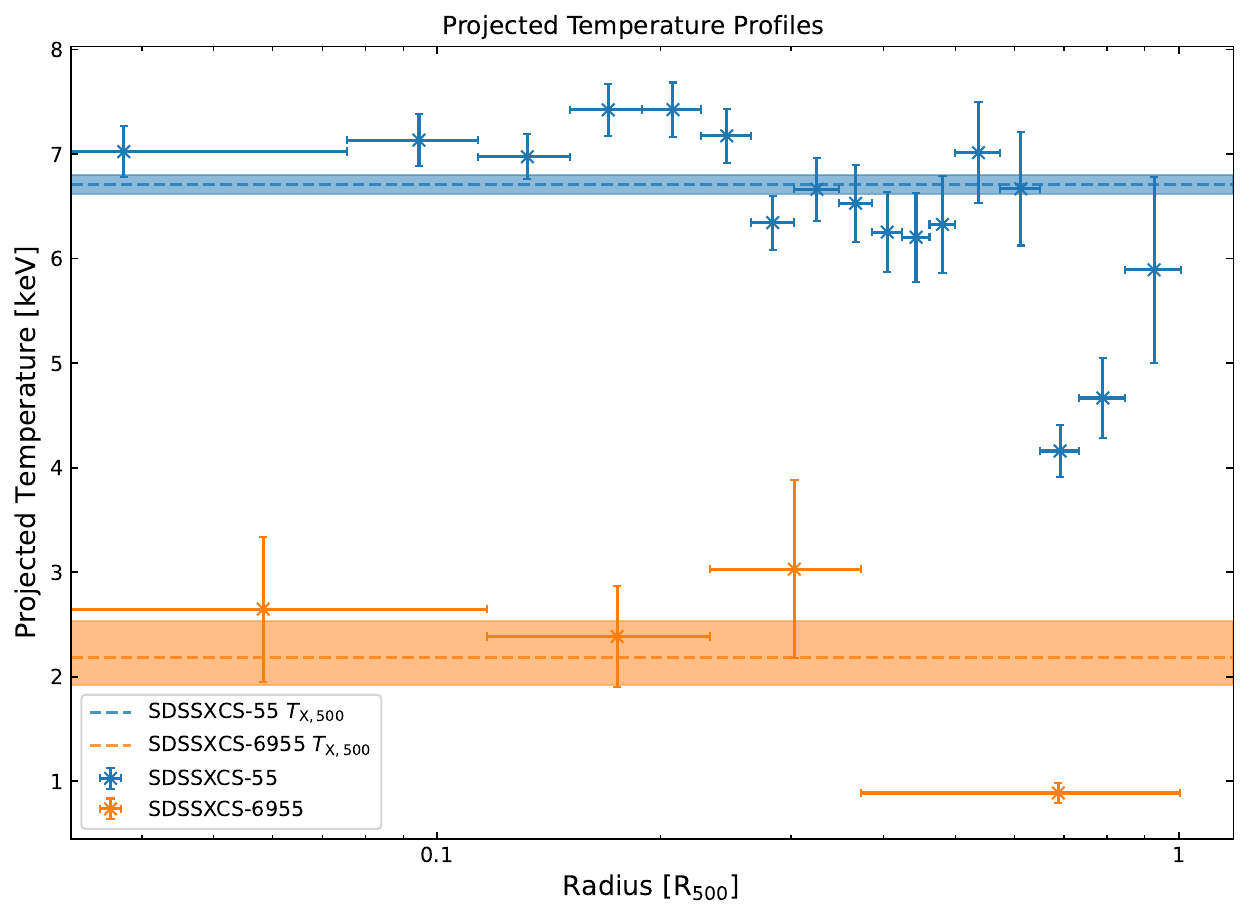}
    \caption[projtemp profs]{Projected temperature profile measurements for the demonstration
    %KR Feb 9 - need to be consistent - demonstration or example?
    clusters, \clustone{} (blue; upper points) and \clusttwo{} (orange; lower points). Temperatures are plotted against the central radii of the annuli they were measured within. The global temperature values measured by \xga{} within $R_{500}$ for each cluster is indicated by horizontal lines, with the shaded region representing 1$\sigma$ uncertainty.}
    \label{fig:projtemp}
\end{figure}

The fits are performed using the $c$-statistic \citep[][]{cash} using a methodology described in detail in \cite[][section 3]{xcsgiles}. Note that not all spectra will yield fitted parameters. There are several quality checks in the \cite{xcsgiles} method that have been replicated in \xga{}. If a given spectrum fails one of those checks, then it is removed from the analysis. If all spectra related to a given analysis region are removed in this way, then no spectral parameter fits will be reported by \xga{}.
Spectral fitting results for our two example clusters are shown in Figure~\ref{fig:projtemp}.
Validation tests of the \xga{} spectral fitting process can be found in Section~\ref{sec:validation}. 

\subsection{Correcting images for PSF distortion}
\label{subsec:psfcorr}

\begin{figure*}
    \centering    
\includegraphics[width=1.0\textwidth]{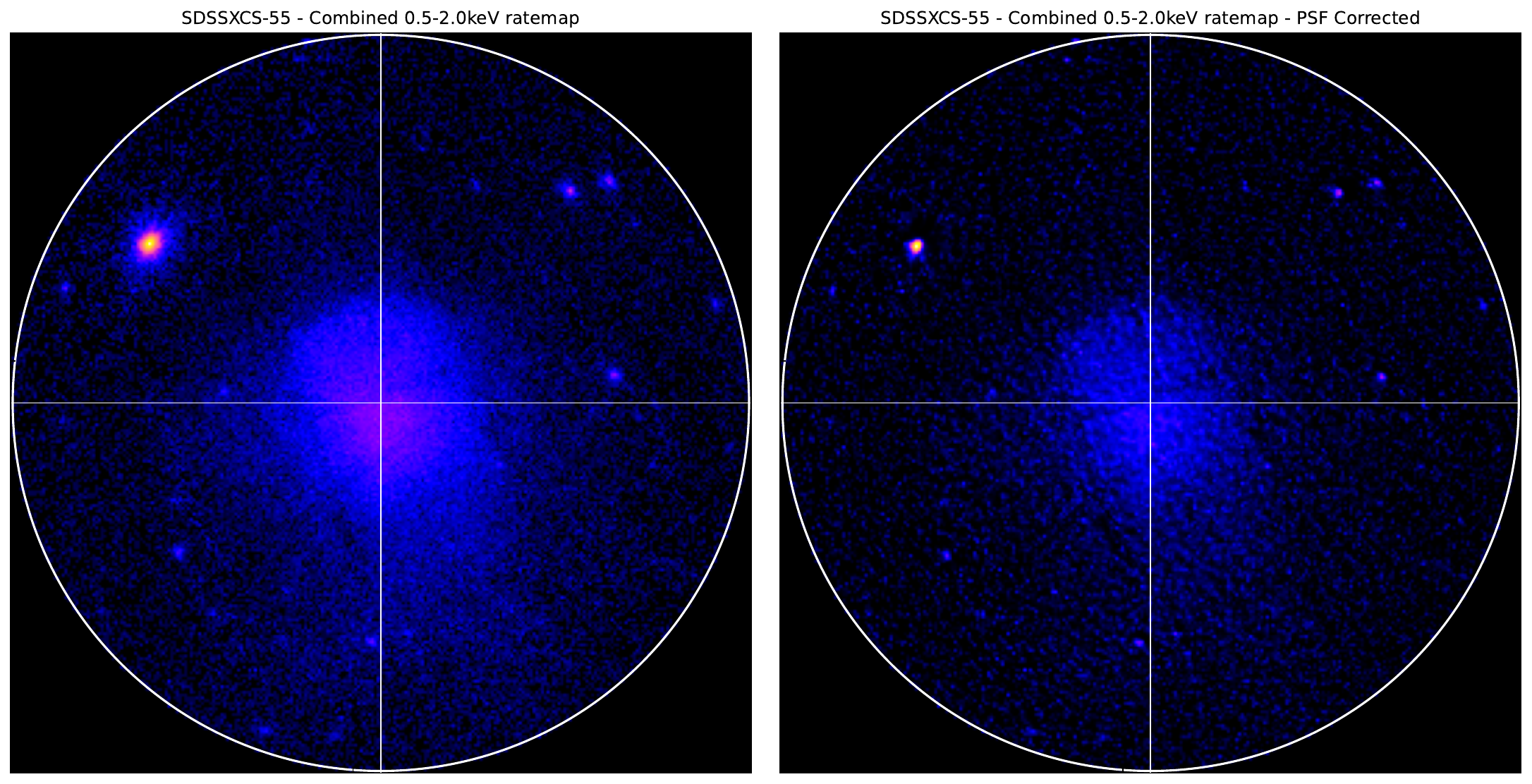}
    \caption[PSF comparison]{
    A demonstration of the image PSF correction capabilities of \xga{}. The white cross-hair indicates the UDC$^{\ref{foot:udc}}$ of \clustone{}. The white circle indicates the $R_{500}$ aperture centered on the UDC$^{\ref{foot:udc}}$. Both images show the $2R_{500}\times2R_{500}$ region. Left: the combined RateMap before PSF correction. 
    Right: After PSF correction. Note, individual camera images were corrected separately, and then stacked.}
    \label{fig:demopsfcorr}
\end{figure*}

The \xmm{} EPIC-PN on-axis PSF has a full width half maximum of ${\sim}12.5\arcsec$ \citep[][]{xmmhandbook}, and the EPIC-MOS1 and MOS2 camera on-axis PSFs have a FWHM of ${\sim}4.3\arcsec$. The PSF of all three EPIC cameras changes size and shape depending on the position on the detector, with the PSF causing stretching of bright sources along the azimuthal direction \citep[e.g.][]{readpsf,xmmhandbook}.
Therefore, \xga{} has been configured to make a PSF correction to \xmm{} images. These corrections are important to our goal of estimating $M_{\rm hydro}$ because the $\rho(r)$ profiles rely on surface brightness maps (see Section~\ref{subsec:SBandEm}). The PSF correction approach deployed in \xga{} uses the Richardson-Lucy algorithm \citep[][]{richardson,lucy} combined with the \texttt{ELLBETA}\footnote{\texttt{ELLBETA} \citep{readpsf} has been implemented in \texttt{SAS} via the \texttt{psfgen} tool.}  XMM PSF model. As the \xmm{} PSFs vary with position, the de-convolution is carried out separately in a user defined  grid across the image. For the analysis here in, we used a $10\times10$ grid. Figure~\ref{fig:demopsfcorr} shows a comparison of the pre and post PSF correction combined count-rate maps for our two example clusters. 

\subsection{Generating surface brightness and emissivity profiles}
\label{subsec:SBandEm}

\begin{figure}
    \centering
    \includegraphics[width=1.0\columnwidth]{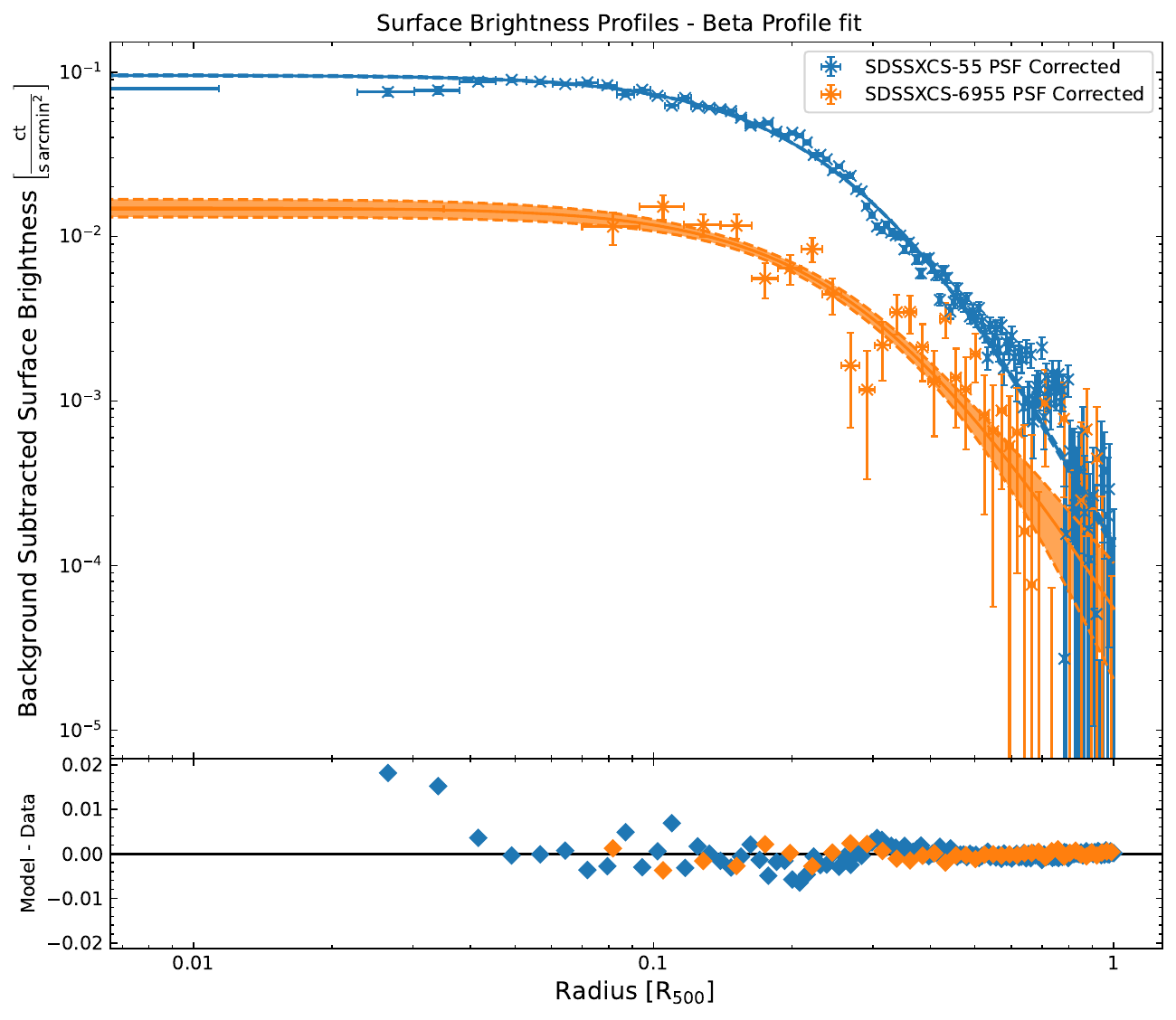}
    \caption[SB profile]{
    Surface brightness profiles (constructed from images in the 0.5-2.0~keV band) generated by \xga{} for \clustone{} (blue; upper points) and \clusttwo{} (orange; lower points), with the spatial binning set to one pixel (4.35\arcsec). Residuals to the fit to a beta model are shown in the bottom panel.}
    \label{fig:sbprof}
\end{figure}

\xga{} can be used to construct radial surface brightness (SB) profiles from combined (Section~\ref{subsec:xgaphot}) PSF corrected (Section~\ref{subsec:psfcorr}) ratemaps. For this, the region (radius and centre) over which the SB is determined is defined by the user as an initiation input to \xga{} (Section~\ref{subsec:xcsdata}). A background region also needs to be defined. In \xga{}, the background region is an annulus. For our example clusters, \clustone{} and \clusttwo{} (and for the analysis presented in Section~\ref{sec:results}), the inner and outer radii were set at 1.05$R_{500}$ and 1.5$R_{500}$ respectively.
Other sources in the respective \xmm{} combined ratemaps were removed prior to profile generation using the automated and  \xga{} generated masks (see Sections~\ref{subsec:xgacontam}, and \ref{subsec:manualdetreg}). For the analyses herein (unless otherwise stated), we used the default \xga{} settings radial bins of width 1~pixel (4.35\arcsec) and an energy range of 0.5-2.0~keV. The uncertainties on the background-subtracted SB profile are calculated by assuming Poisson errors on the counts in each annulus.
Figure~\ref{fig:sbprof} shows cluster surface-brightness profiles for \clustone{} and \clusttwo{}. 

The SB profiles are projected quantities and are in instrument specific units (i.e. detected photons per second per unit area).
The next step toward our goal of measuring a $\rho(r)$ profile is to de-project them and to infer a three-dimensional profile (i.e. emitted photons per second per volume). As the true 3D distribution of ICM electrons is unknown, we have to rely on a model for the de-projection. Fitting to radial profiles using \xga{} is described in Appendix~\ref{app:radmodfit}.  We use the beta model throughout this work, but other models are available in \xga{}. This model takes the form:
\begin{equation}
\label{equ:betamodel}
    S_X(R) =  S_0\left( 1 + \left( \dfrac{R}{R_c} \right)^{2}\right)^{-3\beta + 0.5},
\end{equation}
where $S_{0}$ is a normalisation factor, $R_{c}$ is the radius of the core region and $\beta$ is the slope outside the core region.  Parameter priors used during this work can be found in Table~\ref{tab:models}.  An advantage of the beta model is that there is an analytical solution for the inverse-abel transform which is used to deproject it from a 2D to a 3D profile, making the assumption of spherical symmetry. This results in a 3D radial volume emissivity profile $\epsilon(R)$.

\subsection{Generating density profiles}
\label{subsec:densityProf}

%start paragraph with E(r) and then introduce N to make it clearer

With the emissivity profiles in hand, the next task is to convert them to gas density profiles. For this we use of the definition of the APEC emission model normalisation,

\begin{equation}
    N_{\texttt{APEC}} = \dfrac{10^{-14}}{4\pi(D_{A}(1+z))^2}\int n_e n_p dV,
	\label{eq:apecnorm}
\end{equation}
where $N_{\texttt{APEC}}$ is the normalisation of the \texttt{APEC} plasma emission model (in units of $\rm{cm}^{-5}$), $D_{A}$ is the angular diameter distance to the cluster (in units of cm), $z$ is the redshift of the cluster, $n_{e}$ and $n_{p}$ are the electron and proton number densities (in units of $\rm{cm}^{-3}$). %This will help infer the gas density profile by enabling a conversion between the volume emissivity profile, $\epsilon(R)$, and the density profile.

To make use of Equation~\ref{eq:apecnorm}, we need to calculate a conversion factor, $K_N$, between $N_{\texttt{APEC}}$ and \xmm{} count-rate ($C_r$), where $N_{\texttt{APEC}} = K_N C_r$. To calculate $K_N$, we implement an \xga{} interface to the \xspec{} tool \texttt{FakeIt}, which can be used to simulate spectra given an emission model and an instrument response.  This is used to generate simulated spectra for every camera-plus-observation combination (using ARFs and RMFs extracted at the UDC$^{\ref{foot:udc}}$ during the generation 
of spectra, see Section~\ref{subsec:xgaspecs}). The simulations are performed using an \texttt{APEC} model absorbed with \texttt{tbabs}. The ${\rm n}_{\rm H}$ value for \texttt{tbabs} is set to the \cite{nh} value for that cluster. The spectra are simulated with a normalisation fixed at 1, a fixed global temperature measured for that cluster (within $R_{500}$ for LoCuSS High-\lx{}/SDSSRM-XCS, within 300~kpc for XXL-100-GC), a metallicity of 0.3~$Z_{\odot}$, and the input redshift of the cluster. The simulated spectra are all `observed' for a set exposure time of 10~ks, and are used to measure count-rates in the 0.5-2.0~keV band, which corresponds to the energy range of the ratemaps used to generate SB profiles.  The count-rates for each instrument of each observation are then weighted by the average effective area between 0.5-2.0~keV 
(drawn from the corresponding ARF) and combined into a single conversion factor $K_N$. This conversion factor is suitable for use with emissivity profiles generated from combined ratemaps.

This conversion factor, $K_N$, allows Equation~\ref{eq:apecnorm} to be written in terms of the 3D emissivity $\epsilon(R)={C_{r}(R)}{V}^{-1}$, and the product of the electron and proton number densities $n_{e}n_{p}$ (which we can use to calculate the total gas density), 

\begin{equation}
    n_{e}(R)n_{p}(R) = \dfrac{4\pi K_{N}(D_{A}(1+z))^2\epsilon(R)}{10^{-14}}.
	\label{eq:halfway}
\end{equation}

At this point we assume the ratio of electrons to protons in the intra-cluster medium, given by $n_{e}=R_{ep}\times n_{p}$, which substituted into Equation~\ref{eq:halfway}, gives an expression for $n_{p}$ of
\begin{equation}
    n_{p}(R) = \sqrt{\dfrac{4\pi K_{N}(D_{A}(1+z))^2\epsilon(R)}{R_{ep}10^{-14}}}.
	\label{eq:np}
\end{equation}
We choose to use the solar abundances presented in \cite{andgrev} to calculate this ratio, $R_{ep}=1.199$.
The total gas number density, $n_{g} = n_{e}+n_{p} = (1+R_{ep})n_{p}$, is thus calculated using the expression for $n_{p}$ from Equation~\ref{eq:np}. 

At this point the mean molecular weight $\mu=0.61$ and the atomic mass unit are used to convert number density to mass density. Thus, we calculate gas density from emissivity with quantities that we can measure, or already know;

\begin{equation}
    \rho_{\rm{gas}}(R) = \mu m_{u}(1+R_{ep})\sqrt{\dfrac{4\pi K_{N}\epsilon(R)(D_{A}(1+z))^2}{R_{ep}10^{-14}}}.
	\label{eq:dens}
\end{equation}

\begin{figure}
    \centering \includegraphics[width=1.0\columnwidth]{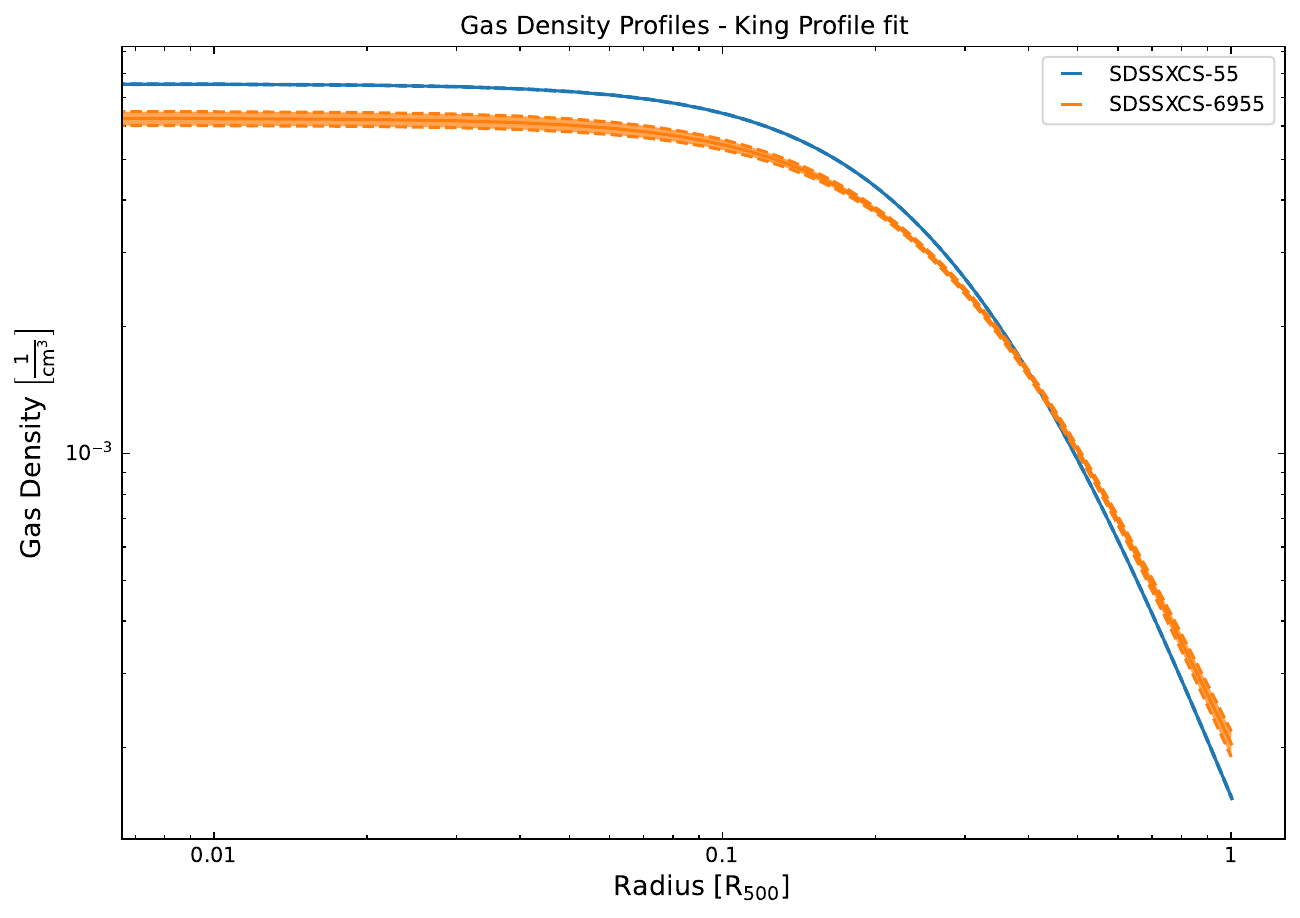}
    \caption[rho profile]{Gas density models generated by \xga{} for \clustone{} (blue) and \clusttwo{} (orange). These are determined from the model fits to surface-brightness profiles shown in Figure~\ref{fig:sbprof}.}
    \label{fig:densprof}
\end{figure}

Once the radial gas density profile is measured using Equation~\ref{eq:dens}, we use the fitting functionality discussed in Appendix~\ref{app:radmodfit} to fit a parametric model to the resulting data points. Density profiles for SDSSXCS-55 and SDSSXCS-6955 are shown in Figure~\ref{fig:densprof}.

% Rolled back this feature - it will be back by the time the reviewers are done I should think
% However, as the density points are themselves generated from a model fit (the double-$\beta$ model in Equ.~\ref{equ:betamodel}, see Table~\ref{tab:models} for further details), and that model has an analytical solution to the inverse-Abel deprojection, we choose to calculate deprojected parameter posterior distributions. This is more computationally efficient than fitting another profile, and better propagates the errors on the surface-brightness profile to the density profile. 

\subsection{Measuring gas masses}
\label{subsec:measgasmass}

Once a gas density profile (Equation~\ref{eq:dens}) has been derived, we can use it to measure total gas masses within given radii. This is achieved through a spherical volume integral, 
\begin{equation}
    M_{\rm{gas}}(<R_{\Delta}) = 4\pi \int_{0}^{R_{\Delta}} \rho_{\rm{gas}}(R)R^2 dR,
	\label{eq:volint}
\end{equation}
and can be used to measure both total gas masses within particular radii (e.g. $R_{500}$), or to create cumulative gas mass profiles.
%(we do not use those in this work).

It is desirable to account for all significant sources of uncertainty in the calculation of gas masses.  One source of uncertainty is that on the overdensity radii within which gas mass is measured, which is not accounted for in our measurement of the density profile. Therefore, we have added an optional mechanism to account for uncertainty in the physical radius (e.g. $R_{500}$) when calculating gas mass. For this, we assume a Gaussian posterior distribution for the radius, with the mean being the published value and the standard deviation being the published error. This radius distribution is sampled along with the model posterior distributions to create a gas mass measurement distribution, with the sampled radius for each combination of sampled model parameters acting as the outer radius within which the integral is evaluated. We have applied this method in Section~\ref{subsec:compgasmasses} where we compare \xga{} and \cite{xxlbaryon} gas mass estimates for the XXL-100-GC sample.  This is due to the fact that the gas masses measured in \cite{xxlbaryon} also account for overdensity radius uncertainty in their analysis.  We also use it in Section~\ref{sec:results} when generating new $M_{\rm{hydro}}$ estimates for SDSSRM-XCS clusters. 

To demonstrate the effect of including the uncertainty on the input radius, the measured gas mass for \clusttwo{} within $R_{500}$ ($669.0^{+62.0}_{-47.9}$ kpc; see Table~\ref{tab:exampclusters}) is $M^{\rm{gas}}_{500}= 1.155^{+0.027}_{-0.027}$ $10^{12}\: \rm{M}_{\odot}$ excluding the uncertainty on $R_{500}$, and $M^{\rm{gas}}_{500}= 1.153^{+0.089}_{-0.095}$ $10^{12}\: \rm{M}_{\odot}$ including the $R_{500}$ uncertainty. We can see that the gas mass uncertainties, when we account for radius uncertainty, are ${\sim}3$ times larger than when we don't.

\subsection{Generating 3D temperature profiles}
\label{subsec:gastempprof}

\begin{figure}
    \centering
    \includegraphics[width=1.0\columnwidth]{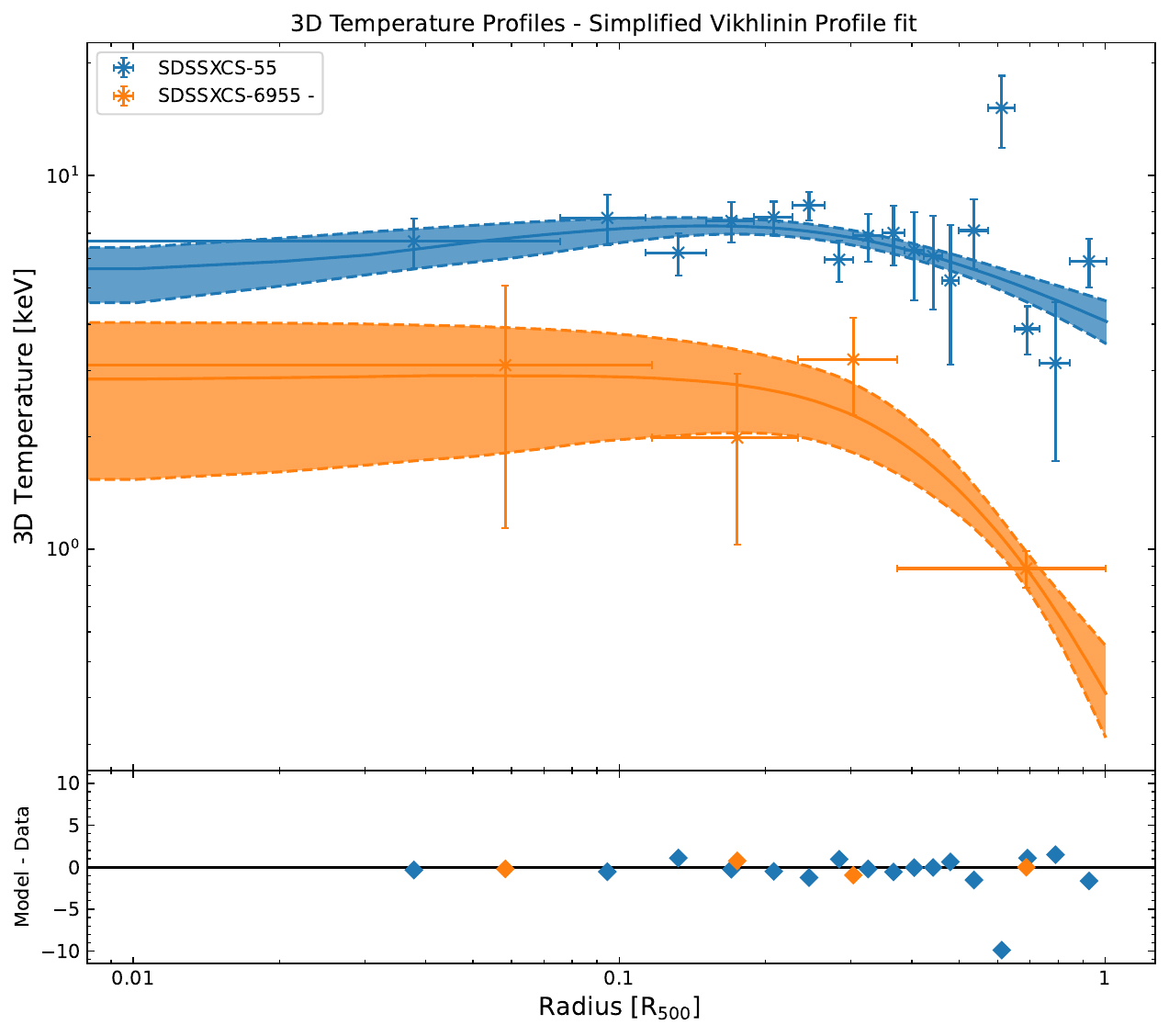}
    \caption[3D T profile]{Three-dimensional temperature profiles, de-projected from the temperature profiles shown in  Figure~\ref{fig:projtemp}, for \clustone{} (blue points) and \clusttwo{} (orange points).  Fits to the temperature profiles using the simplified Vikhlinin model in Equation~\ref{equ:simp_temp_prof}, using the process outlined in Appendix~\ref{app:radmodfit}, are given by the blue (\clustone{}) and orange (\clustone{}) solid lines.  For each fit, the shaded regions, bounded by the dashed lines, represent the 1$\sigma$ uncertainty. Residuals to the fit are shown in the bottom panel.} 
    \label{fig:3dtempprof}
\end{figure}

The second property we must measure to calculate a cluster's hydrostatic mass is the three-dimensional temperature profile. The radial spectral analysis described in Section~\ref{subsec:xgaspecs} results in a set of projected temperature measurements (one for each of the $N$ radial bins).
Each projected temperature is a weighted combination of the temperatures in the three-dimensional shells of the cluster that the annulus intersects with. To recover the 3D distribution, $T(r)$, we opt to use the `onion-peeling' method \citep[see descriptions in e.g.,][]{onion1,onion2}. First, the volume intersections between the analysis annuli (projected back into the sky) and the spherical shells which we separate each cluster into (defined as having the same radii as the annuli) must be calculated. The volume intersections can be calculated  \citep[see Appendix A of][]{volintersec} as, 

\begin{equation}
\begin{split}
    \bm{V}_{\rm{int}} = \frac{4\pi}{3}\big[&max\{0, (\bm{R}_{\rm{so}}^{2} -  \bm{R}_{\rm{ao}}^{2})^{\frac{3}{2}}\} - max\{0, (\bm{R}_{\rm{so}}^{2} - \bm{R}_{\rm{ao}}^{2})^{\frac{3}{2}}\} +\\ & max\{0, (\bm{R}_{\rm{si}}^{2} - \bm{R}_{\rm{ao}}^{2})^{\frac{3}{2}}\} - max\{0, (\bm{R}_{\rm{si}}^{2} - \bm{R}_{\rm{ai}}^{2})^{\frac{3}{2}}\} \big],
\end{split}
\label{eq:intersec}
\end{equation}
where $\bm{R}_{\rm{so}}$ is the matrix of shell outer radii, $\bm{R}_{\rm{ao}}$ is the matrix of annulus outer radii, $\bm{R}_{\rm{si}}$ is the matrix of shell inner radii, and $\bm{R}_{\rm{ai}}$ is the matrix of annulus inner radii.  
The $\bm{V}_{\rm{int}}$ matrix is a two dimensional matrix describing volume intersections between all combinations of annuli and three-dimensional shells.

Second, an emission measure profile is calculated using the APEC Normalisation 1D profile produced during the spectral fitting of the set of annular spectra. This calculation uses Equation~\ref{eq:apecnorm} and rearranges to solve for the integral. As the projected temperature measured within a given annulus is a weighted combination of the shell temperatures intersected by that annulus, we can infer the temperature of a shell \citep[for past usage see][]{onion2} with 

\begin{equation}
    \bm{T}_{\rm{shell}} = \dfrac{(\bm{V}_{\rm{int}}^{\rm{T}})^{-1} \times \bm{T}_{\rm{annulus}}\rm{\textbf{EM}}}{(\bm{V}_{\rm{int}}^{\rm{T}})^{-1} \times \rm{\textbf{EM}}}.
	\label{eq:tempdeproj}
\end{equation}

$\bm{T}_{\rm{shell}}$ is the matrix of spherical shell temperatures that we aim to calculate, $\bm{V}_{\rm{int}}$ is the matrix of volume intersections between all combinations of annuli and shells (see Equation~\ref{eq:intersec}), $\rm{\textbf{EM}}$ is the emission measure matrix, and $\bm{T}_{\rm{annulus}}$ is the projected temperature matrix; $\times$ represents the matrix product.
We propagate the uncertainties from the projected temperature and emission measure profiles by generating 10000 realisations of each profile, assuming Gaussian errors on each data point. We then use the profile realisations to calculate 10000 instances of the deprojected temperature profile. These distributions are used to measure 90\% confidence limits on each deprojected temperature data point. 

Once three-dimensional, de-projected, gas temperature profiles have been measured, we use the methods discussed in Section~\ref{app:radmodfit} to model the profiles with a simplified version of the \cite{vikhlininmass} temperature model as detailed in \cite{onion2}.  The use of the simplified model allows for the modelling of temperature profiles with fewer temperature bins (due to the use of less free parameters). This temperature profile takes the form:

\begin{equation}
\label{equ:simp_temp_prof}
    T_{3D}(r) = \dfrac{T_{0}\left(\left(\frac{r}{r_{\rm cool}}\right) + \frac{T_{\rm min}}{T_{0}} \right)\left( \frac{r^{2}}{r_{t}^{2}} + 1 \right)^{-\frac{c}{2}}}{\left(\frac{r}{r_{\rm cool}}\right)^{a_{\rm cool}} + 1},   
\end{equation} 
where $T_{0}$ is a normalisation factor, $T_{\rm min}$ is the minimum temperature, $r_{\rm cool}$ is the radius of the cool central region, $a_{\rm cool}$ is the slope of the cool region out to a radius $r_{\rm cool}$, c is the slope at large radii occurring at a transition radius $r_{t}$. Parameter priors can be found in Table~\ref{tab:models}.
Figure~\ref{fig:3dtempprof} shows three-dimensional temperature profiles for clusters \clustone{} and \clusttwo{}, along with the corresponding 1$\sigma$ uncertainty (given by the shaded regions).

\subsection{Generating hydrostatic mass profiles}
\label{subsec:hydromassprof}

\begin{figure}
    \centering
    \includegraphics[width=1.0\columnwidth]{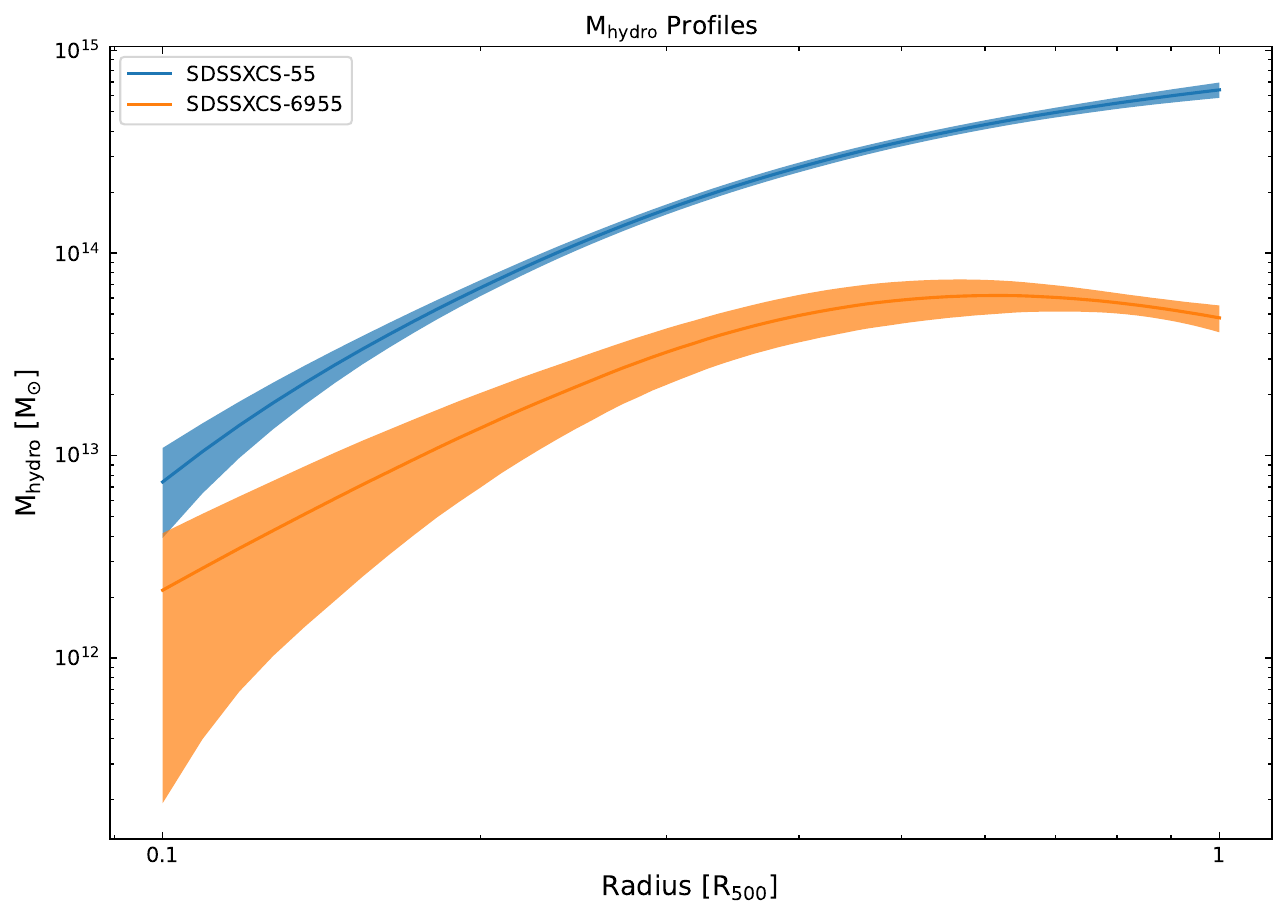}
    \caption[Hmas profile]{An example hydrostatic mass profile generated for \clustone{}. This profile was generated using the model fits to the density profile shown in Figure~\ref{fig:densprof}, and the temperature profile shown in Figure~\ref{fig:3dtempprof}.}
    \label{fig:hymprof}
\end{figure}

Once we have measured a 3D radial gas density profile (details in Section~\ref{subsec:densityProf}) and a 3D radial temperature profile (details in Section~\ref{subsec:gastempprof}), we can use Equation~\ref{eq:hydromass} to measure hydrostatic masses. As such, \xga{} creates hydrostatic mass profiles as a function of radius (e.g. Figure~\ref{fig:hymprof}).
The hydrostatic mass equation not only involves the temperature and density profile values at each radius for which a enclosed mass is measured, but also the derivatives of those profiles with respect to radius. The king profile and temperature profiles both have analytical first-derivatives, and as such these are used to calculate the slope at a given radius, rather than a numerical approximation. 

\xga{} models can return posterior distributions of the value of the model (or model slope) at a given radius, rather than a single value. This is based on drawing randomly from the parameter posterior distributions found from the fitting process described in Section~\ref{app:radmodfit}. When calculating a hydrostatic mass at a specified radius, the temperature and density parametric models generate 10000 realisations of themselves; these realisations are then used to retrieve absolute values of $T_{\rm{3D}}$ and $\rho_{\rm{g}}$ at the specified radius, as well as $\frac{dT_{\rm{3D}}}{dr}$ and $\frac{d\rho_{\rm{g}}}{dr}$ values. As such a distribution of hydrostatic mass measurements is created for the given radius, and 90\% confidence limits are calculated.

We also implement an optional method of propagating errors on the chosen radius, to help account for uncertainties on the overdensity radii which masses are commonly calculated within. This is akin to the optional step implemented for the calculation of \xga{} gas masses at the end of Section~\ref{subsec:measgasmass}. When an uncertainty is provided along with a radius, we assume a Gaussian distribution and draw 10000 random radii. Then, when the profile realisations are generated and a distribution of hydrostatic masses are measured, the randomly drawn radii are used rather than a fixed value.  Distributions are generated for absolute values and derivatives can all take radius uncertainties into account. This is not used in our comparisons to LoCuSS $M_{\rm{hydro}}$ measurements in Section~\ref{subsec:mhydrocomp}, as radius uncertainties were not published by \cite{locusshydro}. We do include radius uncertainties in our calculation of SDSSRM-XCS cluster masses presented in Section~\ref{sec:results}. Mass profiles for the clusters \clustone{} and \clusttwo{}, along with the corresponding 1$\sigma$ uncertainty (given by the shaded regions), are shown in Figure~\ref{fig:hymprof}.

\section{Validation Tests}
\label{sec:validation}

We have tested the validity of the \xga{} approach described in Section~\ref{sec:methodology} by comparing the \xga{} outputs to those presented in the literature. For this we have used three different cluster samples. The samples used for validation are  described in Section~\ref{subsec:samples}. In Sections~\ref{subsubsec:txtests}, \ref{subsec:compgasmasses}, \ref{subsec:mhydrocomp} we compare the \xga{} measurements of $T_{\rm X}$, $M_{\rm{gas}}$, $M_{\rm{hydro}}$ respectively to literature values. 

\begin{table*}
\centering
\caption[]{{Summaries of the literature galaxy cluster samples used in this work (see Section~\ref{sec:validation}) to validate the \xga{} methodology.}
\label{tab:samplesummary}
}
\vspace{1mm}

\begin{tabular}{|ll| p{0.4\linewidth} | p{0.30\linewidth}}
% \begin{tabular}{l|lc}
\hline
\hline
 Sample Name & N$_{\rm{CL}}$, $z$ & Brief description\\ 
\hline
\hline
SDSSRM-XCS & 
150, $0.1{<}z{<}0.35$ 
& 
SDSS redMaPPer clusters \citep[][]{redmappersdss} with available \xmm{} observations. Section~\ref{sec:validation} presents a comparison to results presented in \cite{xcsgiles}.
\\
\hline
XXL-100-GC & 99, {$0.05<z<0.9$} & Drawn from a sample of the 100  X-ray brightest clusters in the XXL survey regions \citep[][]{xxl1,gc100}.
Section~\ref{sec:validation} presents a comparison to results presented in \cite{xxllt} and \cite{xxlbaryon}.
\\
\hline
LoCuSS High-$L_{\rm{X}}$ (a) & 33, $0.15<z<0.3$ & Drawn from clusters detected in the the ROSAT All Sky Survey \citep{rass_ebeling}.  The sample contains 50 clusters satisfying the following conditions; ${\rm n}_{\rm H}<7\times10^{20}$~cm$^{2}$, $-25^{\circ} < \delta < 65^{\circ}$ and $L_{X}[0.1 - 2.4~{\rm keV}]E(z)^{-2.7} \geq 4.2\times10^{44}$ erg s$^{-1}$.  All clusters have subsequent observations by \xmm{} or \chandra{}. Section~\ref{sec:validation} presents a comparison to 33 clusters with \xmm{} derived results presented in \cite{locusshydro}.
\\
LoCuSS High-$L_{\rm{X}}$ (b) & 32, $0.15<z<0.3$ & As above, but for clusters satisfying $L_{X}[0.1 - 2.4~{\rm keV}]E(z)^{-1} \geq 4.4\times10^{44}$ erg s$^{-1}$ between $0.15 < z < 0.24$, and $L_{X}[0.1 - 2.4~{\rm keV}]E(z)^{-1} \geq 7.0\times10^{44}$ erg s$^{-1}$ between $0.24 < z < 0.3$.  Resulting in a sample of 41 clusters. Section~\ref{sec:validation} presents a comparison to 32 clusters with \xmm{} derived results presented in \cite{locusstemps}.
\\
\hline
\end{tabular}
\end{table*}

\subsection{Validation Samples}
\label{subsec:samples}

The validation samples are SDSSRM-XCS \citep{xcsgiles}, XXL-100-GC \citep[][]{gc100}, and LoCuSS High-$L_{\rm{X}}$ \citep{locusshydro}. The properties of the validation samples are summarised in Table~\ref{tab:samplesummary}. We note that although both \xmm{} and \chandra{} based measurements are presented in \cite{locusshydro},
%comment: and in Mulroy? KR
we only make comparisons here with \xmm{}, given the known discrepancy between \xmm{} and \chandra{} derived  temperature estimates \cite[e.g.,][]{schellenbergertcal}. Excluding duplicates,
a total of \totaluniqclustersvaltests{} clusters have been used in the validation tests presented below. Duplicate clusters were identified by cross-matching the samples; to be considered a match, two sources needed to be within a projected distance of 500~kpc (at the cluster redshift) and with $\lvert\Delta z\rvert \leq 0.05$. For each test, we adjusted the \xga{} initiation values (Section~\ref{subsec:xcsdata}) to follow those used in the published works as closely as possible (see Sections~\ref{subsubsec:SDSSRMsim}, \ref{subsubsec:XXLsim}, \ref{subsubsec:LoCuSSsim}).
A companion GitHub repository (see Appendix~\ref{app:companionrepo} for a summary of the structure and contents) contains the exact sample files used in this section. 

\subsubsection{SDSSRM-XCS}
\label{subsubsec:SDSSRMsim}

We have used \xga{} to re-analyse a sample of 150 clusters presented in \citet[][hereafter G22]{xcsgiles}. These clusters are referred to as the SDSSRM-XCS$T_{\rm X,vol}$ sub-sample in G22 (see Table 2 therein), but hereafter will be referred to as SDSSRM-XCS for simplicity. The SDSSRM-XCS clusters in the sample were originally selected from SDSS photometry using the redMaPPer (RM) algorithm \citep[][]{redmappersdss}. The 150 SDSSRM-XCS clusters represent the subset of the original $\simeq$66,000 SDSSRM cluster catalogue that meet the following criteria: they lie within the footprint of the the \xmm{} archive, were successfully processed by the XCS imaging and spectroscopic pipelines ($\Delta T_{\rm X}<25\%$), 
%KR Feb 9 - has delta T been explained in \S2 ? ref back to it?
and have redshifts in the range $0.1<z<0.35$. 

The following elements are in common between G22 and our analysis of the SDSSRM-XCS clusters:

\begin{itemize}[leftmargin=*, labelindent=2pt, itemsep=1ex]
\item[$\bullet$] The cosmological model; flat $\Lambda$CDM, assuming $\Omega_{\rm{M}}$=0.3, $\Omega_{\Lambda}$=0.7, and $\rm{H}_{0}$=70 km s$^{-1}$ Mpc$^{-1}$.  
% \item[$\bullet$] The input \xmm{} observations. Even though \xmm{} observations are added to the public archive on a regular basis, {\color{red} no new, compared to G22, observations were used in the validation tests.} In total {\color{red}\sdssrmobsids} were used.
\item[$\bullet$] The \xmm{} data reduction, image generation, and source detection
%comment: and calibration? KR
(Section~\ref{subsec:xcsdata}).
%comment: correct? needs better explanation KR
This was performed outside of \xga{}, exactly following the G22 approach. 
%comment: is the same? is it? did we use all observations? KR
This meant that the automatically masked regions (Section~\ref{subsec:xgacontam})  were the same in both analyses. 
\item[$\bullet$] The UDCs$^{\ref{foot:udc}}$, redshifts, $R_{500}$, and $R_{2500}$ values. Therefore, the source apertures, and the background annuli (either 1.05-1.5$R_{500}$ or 2-3$R_{2500}$) were the same (i.e. size, shape, and location) during the global spectral fits (Section~\ref{subsec:xgaspecs}), and construction of surface brightness profiles (Section~\ref{subsec:SBandEm}).
\item[$\bullet$] The model used during the \xspec{} fits. Both analyses used an absorbed APEC plasma model with the column density, redshift, and metal abundance fixed during the fit. 
\end{itemize}

Differences include:

\begin{itemize}[leftmargin=*, labelindent=2pt, itemsep=1ex]
\item[$\bullet$] The choice of manually adjusted source masks and excluded \xmm{} 
observations (Section~\ref{subsec:manualdetreg}).
\item[$\bullet$] Different versions of certain software packages; G22 used SAS v14.0.0 and \xspec{} v12.10.1f, whereas this work uses SAS {v18.0.0} and  \xspec{} {v12.11.0}.
\item[$\bullet$] The pipeline used for the global spectral analysis (there was no radial spectral analysis G22). Herein we use \xga{}, whereas G22 used XCS3P. One important difference between the two pipelines is the treatment of \xmm{} sub-exposures. Some \xmm{} observations contain multiple sub-exposures by the same instruments. The analyses performed by \xga{} only make use of the longest individual sub-exposure for a particular instrument of a particular observation, whereas XCS3P makes use of all of the sub-exposures. 
\end{itemize}

\subsubsection{XXL-100-GC}
\label{subsubsec:XXLsim}

We used \xga{} to re-analyse 99 of the 100 clusters first described in \cite{gc100}. 
 The cluster XLSSC-504 was excluded to be consistent with the \cite{xxllt} study.  A further two clusters were excluded as they did not successfully pass \xga{} data quality checks:
\texttt{backscale} errors were encountered during spectral generation for XLSSC-11, 
and, for  XLSSC-527, no observation fulfils the \xga{} criterion that the 300~kpc coverage (Section~\ref{subsec:xcsdata}) fraction should be $>70\%$. We compare the \xga{} derived global temperature measurements of the remaining 97 to those presented in 
\cite{xxllt}\footnote{\href{http://vizier.u-strasbg.fr/viz-bin/VizieR-3?-source=IX/49/xxl100gc}{http://vizier.u-strasbg.fr/viz-bin/VizieR-3?-source=IX/49/xxl100gc}}, and the \xga{} derived gas mass measurements to those presented in \cite{xxlbaryon}\footnote{The uncertainties on $R_{500}$ values used in Section~\ref{subsec:compgasmasses} are retrieved directly from \cite{xxlbaryon}.}. 

Our analysis of the XXL-100-GC clusters included these elements in common with the published works:

\begin{itemize}[leftmargin=*, labelindent=2pt, itemsep=1ex]
\item[$\bullet$] The cosmological model: $\Omega_{\rm{M}}$=0.28, $\Omega_{\Lambda}$=0.72, and $H_{0}$=70 km s$^{-1}$ Mpc$^{-1}$, the WMAP9 results \citep[][]{wmap9cosmo}.
\item[$\bullet$] The UDCs$^{\ref{foot:udc}}$, redshifts, and analysis regions ($r=300$~kpc) of the clusters. Therefore, the source apertures were the same (i.e. size, shape, and location) during the global spectral fits (Section~\ref{subsec:xgaspecs}), and construction of surface brightness (SB) profiles (Section~\ref{subsec:SBandEm}) of both analyses. 

\item[$\bullet$] The model used during the \xspec{} fits was the same. Both analyses used an absorbed APEC plasma model with the column density, redshift, and metal abundance fixed (at 0.3~$Z_{\odot}$, using  \cite{andgrev} abundance tables) during the fit. 
%\item[$\bullet$] A fixed metallicity of 0.3~$Z_{\odot}$, and the \cite{andgrev} abundance tables.
\item [$\bullet$] \xspec{} fitting within an energy range of 0.4-7.0~keV.
\item [$\bullet$] Surface brightness profiles were derived from images generated in the 0.5–2.0keV energy band.
\item [$\bullet$] Surface brightness profiles are generated out to $1.2R_{500}$.
\item[$\bullet$] The global temperature value from \citet[][$T_{\rm{X}}^{300\rm{kpc}}$]{xxllt} was used for the conversion from SB to emission measure (see Section~\ref{subsec:SBandEm}).
\item [$\bullet$] $M_{\rm gas}$ values, and errors, were estimated within R$_{500}$ taken from \cite{xxlbaryon}.  Additionally, $M_{\rm gas}$ uncertainties include the error on R$_{500}$ as performed in \cite{xxlbaryon}, see Section~\ref{subsec:measgasmass}.

\end{itemize}

Differences include:

\begin{itemize}[leftmargin=*, labelindent=2pt, itemsep=1ex]
\item[$\bullet$] The input \xmm{} observations. Our analysis includes some \xmm{} observations that entered the archive after the publication of \cite{xxllt}. 
%In total {\color{red}\xxlobsids} were used.
\item[$\bullet$] 
The approach to background subtraction for the global spectral analyses.  We used a local, in field, subtraction technique with an annulus of width $1.05-1.5R_{500}$ centered on the UDC$^{\ref{foot:udc}}$. By comparison, \citet[][]{xxllt}, used either: {\em (i)} an annulus centered on the aimpoint of the {\em XMM} observation, with a width determined by the diameter of the analysis region of the galaxy cluster \citep[see Figure~1 of][]{xxllt}; or {\em (ii)} an annulus centered on the cluster centroid with an inner radius equal to the extent of the cluster emission and an outer radius a factor 2$\times$ the inner radius.  
\item[$\bullet$] The approach to background subtraction used during the generation of surface brightness (and thus emission measure) profiles. Whereas we used a simple technique (see above), \cite{{xxlbaryon}} used a model for the non-X-ray background and a spatial fit for the X-ray background (see section 2.2 and 2.3 in \cite{{xxlbaryon}} for more detail).
\item[$\bullet$] The choice of manually adjusted source masks and excluded \xmm{} 
observations (Section~\ref{subsec:manualdetreg}).
\item[$\bullet$] Different versions of certain software packages.  
\citet[][]{gc100} used SAS v10.0.2, whereas we use SAS v18.0.0. \cite{xxllt} used \xspec{} v12.8.1i, whereas we use v12.11.0.
\item[$\bullet$] 
%\item[$\bullet$] The \xmm{} data reduction, image generation, and source detection. Whereas herein we followed the methods described in G22, the XXL results are based on methods presented in \citet[][]{gc100}.
The pipelines used to generate the various analysis products required to obtain $T_{\rm X}$ and $M_{\rm gas}(<r)$ estimates. Herein, we use \xga{} (Section~\ref{sec:methodology}), which differs in several ways to those used in \cite{gc100}, \cite{xxllt}, and/or \cite{xxlbaryon}.

\end{itemize}

\subsubsection{LoCuSS High-$L_{\rm{X}}$ (a)}
\label{subsubsec:LoCuSSsim}

We have used \xga{} to analyse the \LOCUSSTxvaluesorginal{} LoCuSS clusters presented in \cite{locusshydro}, for which their gas and hydrostatic masses are determined using only
%KR Feb 9 is "only" correct? Aren't there clusters that have both XMM and Chandra?
{\em XMM} observations.  These clusters form the first LoCuSS validation sample, referred to as LoCuSS High-$L_{\rm{X}}$ (a).  The full sample definition is given in Table~\ref{tab:samplesummary}. We note that we do not measure any results for `RXCJ1212.3-1816', as its sole \xmm{} observation (0652010201) was excluded during the inspection detailed in Section~\ref{subsec:manualdetreg} due to residual flaring. 
Our analysis of the LoCuSS High-$L_{\rm{X}}$ clusters included these elements in common with the published works:

\begin{itemize}[leftmargin=*, labelindent=2pt, itemsep=1ex]
\item[$\bullet$]  The cosmological model: flat $\Lambda$CDM, assuming $\Omega_{\rm{M}}$=0.3, $\Omega_{\Lambda}$=0.7, and $\rm{H}_{0}$=70 km s$^{-1}$ Mpc$^{-1}$. 
\item[$\bullet$] 
The UDCs$^{\ref{foot:udc}}$ (i.e. choice of cluster centroid locations) and redshifts values were taken from \cite{locusshydro}.  To compare gas and hydrostatic masses, we used the R$_{500}$ values given in \cite{locusshydro}. Therefore, the source apertures were the same (i.e. size, shape, and location) in the relevant aspects (global spectral fits, Section~\ref{subsec:xgaspecs}) of both analyses.
\item[$\bullet$] The model used during the \xspec{} fits was the same. The LoCuSS High-$L_{\rm{X}}$ (a) analyses used an absorbed APEC plasma model with the column density and redshift fixed during the fit. The metallicity is left free to vary.  One small difference is that \cite{locusshydro} uses the {\tt wabs} model (rather than {\tt tbabs}) to account for absorption. 
\item We perform \xspec{} fits within an energy range of 0.7-10.0~keV.
%\item[$\bullet$] Metallicity is left free to vary during \xspec{} fits. 
\item [$\bullet$] Surface brightness profiles were derived from images generated in the 0.5–2.5~keV energy band.
\item [$\bullet$] Temperature profiles were constructed with a minimum number of background subtracted counts per annulus of 3000.
%comment: : mention core excised KR
\item [$\bullet$] Global temperature values were measured from core-excised (0.15-1$R_{500}$) spectra.
\item Errors on the gas masses do not include overdensity radius uncertainties (see Sect.~\ref{subsec:measgasmass}) as \cite{locusshydro} did not use this approach in their analysis.

\end{itemize}

Differences include:

\begin{itemize}[leftmargin=*, labelindent=2pt, itemsep=1ex]
\item[$\bullet$] The input \xmm{} observations. Our analysis includes \xmm{} observations that entered the archive after the publication of \cite{locusshydro}. 
\item[$\bullet$] The choice of manually adjusted source masks and excluded \xmm{} 
observations (Section~\ref{subsec:manualdetreg}).
\item[$\bullet$]  The approach to background subtraction. Whereas we used a simple, in field, subtraction technique with an annulus of width $1.05-1.5R_{500}$ (1.05-1.5~$\theta_{\rm{outer}}'$) centered on the UDC$^{\ref{foot:udc}}$, for the global spectral and surface brightness (radial spectral) analyses, \citet[][]{locusshydro}, used a spectral modelling approach (see section 3.3.1 and 3.3.3 in \cite{{locusshydro}} for more detail).
\item[$\bullet$] Different versions of certain software packages; the \cite{locusshydro} work uses SAS v11.0.0, we use SAS v18.0.0. It is not stated which version of \xspec{} was used by \cite{locusshydro}, but it will be a considerably older version than we use (v12.11.0).
\item[$\bullet$]  The pipelines used to generate the various analysis products required to obtain $M_{\rm gas}(<r)$ and $M_{\rm hydro}(<r)$ estimates. Herein, we use \xga{} (Section~\ref{sec:methodology}), which differs in several ways to those used in \cite{locusshydro}.
\item[$\bullet$] The equations to model the density and 3D temperature profiles differ between those used in this work (Equation~\ref{eq:king} and Equation~\ref{eq:simpvikhtemp} for density and temperature respectively) and the form used in \citet[][see their Equation~4 and 5 for density and temperature respectively]{locusshydro}.
\end{itemize}

\subsubsection{LoCuSS High-$L_{\rm{X}}$ (b)}

A modified version of the LoCuSS sample selection is given in \cite{locusstemps}, for which 32 clusters were analysed using only {\em XMM} observations\footnote{The data tables in \cite{locusstemps} did not specific which values were derived from \xmm{} and which from \chandra{}. This information was provided by G. Smith, priv. comm.} to provide global temperature measurements. These clusters form the second LoCuSS validation sample, referred to as LoCuSS High-$L_{\rm{X}}$ (b).  The full sample definition is given in Table~\ref{tab:samplesummary}.  As the \cite{locusstemps} is based largely upon the same analysis as \cite{locusshydro}, many of the similarities and differences are the same as those given in Section~\ref{subsubsec:LoCuSSsim}.  However, those specific to LoCuSS High-$L_{\rm{X}}$ (b) are given below.  The elements in common include:
\begin{itemize}[leftmargin=*, labelindent=2pt, itemsep=1ex]
    \item[$\bullet$] To compare global temperatures, we used the same R$_{500}$ value as that used in \cite{locusstemps}.  The R$_{500}$\footnote{These values were provided by G. Smith via priv. comm.} values were estimated from the weak lensing analysis outlined in \cite{locusswl}. 
\end{itemize}

Differences include:
\begin{itemize}[leftmargin=*, labelindent=2pt, itemsep=1ex]
\item[$\bullet$] Specific information regarding the background subtraction methods used to derive the global $T_{\rm X}$ values provided in \cite{locusstemps} is not given, but we assume the same (spectral modelling) approach was used as that detailed in \cite{locusshydro}.
\end{itemize}

\subsection{Validation of derived properties}

\begin{table}           
\begin{center}
\caption[]{{\small Best fit parameters of fixed-slope power-law models fit to comparisons between original published properties of the validation samples (see Section~\ref{subsec:samples}), and the \xga{} derived properties. The first column indicates the property compared, the second the particular sample (either SDSSRM-XCS, XXL, or LoCuSS High-\lx{}(a) and (b)), the third and forth columns gives the best-fit normalisation and intrinsic scatter (with uncertainties) respectively, and the last column links to the relevant figure.}\label{tab:compfit}}
\vspace{1mm}
\begin{tabular}{ccccc}
\hline
\hline
 Property & Sample & Norm & Scatter & Figure \\
 & & $A$ & $\sigma_{\rm int}$ & \\ \hline
\hline
\vspace{1mm}
$T_{\rm{X}}^{2500}$ & SDSSRM-XCS & 0.99$\pm$0.01 & 0.02$\pm$0.01 & \ref{fig:xcstempcomp}(a) \\
\vspace{1mm}
$T_{\rm{X}}^{500}$ & SDSSRM-XCS & 1.00$\pm$0.01 & 0.04$\pm$0.01 & \ref{fig:xcstempcomp}(b) \\
\vspace{1mm}
$T_{\rm{X}}^{\rm 300kpc}$ & XXL-100-GC & 0.99$\pm$0.01 & 0.04$\pm$0.01 & \ref{fig:xxltxcomp} \\
\vspace{1mm}
$T_{\rm{X}}^{\rm 500ce}$ & LoCuSS High-\lx{}(b) & 1.01$\pm$0.01 & 0.04$\pm$0.01 & \ref{fig:locusstxcomp} \\

\hline
\vspace{1mm}
$M_{\rm{gas}}^{500}$ & XXL-100-GC & 0.89$\pm$0.06 & 0.59$\pm$0.05 & \ref{fig:compxxlgm} \\
\vspace{1mm}
$M_{\rm{gas}}^{2500}$ & LoCuSS High-\lx{}(a) & 0.97$\pm$0.02 & 0.08$\pm$0.01 & \ref{fig:complocussgm}(a) \\
\vspace{1mm}
$M_{\rm{gas}}^{500}$ & LoCuSS High-\lx{}(a) & 0.93$\pm$0.02 & 0.13$\pm$0.02 & \ref{fig:complocussgm}(b) \\
\hline
\vspace{1mm}
$M_{\rm{hy}}^{2500}$ & LoCuSS High-\lx{}(a) & 1.10$\pm$0.03 & 0.08$\pm$0.03 & \ref{fig:complocusshym}(b) \\
$M_{\rm{hy}}^{500}$ & LoCuSS High-\lx{}(a) & 1.19$\pm$0.07 & 0.24$\pm$0.05 & \ref{fig:complocusshym}(b) \\
\hline
\end{tabular}
\end{center}
\end{table}

The validation of the derived properties takes the form of one-to-one (1:1) comparisons to the validation sample described above, and are quantified by fitting a power-law with the slope fixed at unity. The fits were performed in log space using the R package LInear Regression in Astronomy\citep[{\sc lira}\footnote{\href{https://cran.r-project.org/web/packages/lira/index.html}{LInear Regression in Astronomy}}, ][]{softlira}, fully described in \cite{LIRA}. The validations are visualised via 1:1 plots for each property compared, and in each case the best-fit is given by a light-blue solid line and the 68\% confidence interval by the light-blue shaded region.  The 1, 2 and 3-$\sigma$ intrinsic scatter is given by the grey shaded regions.  A summary of fit results can be found in Table~\ref{tab:compfit}. In the following sections we discuss the comparisons of each property for each sample in more detail.  The results can be reproduced using public \href{https://github.com/DavidT3/XCS-Mass-Paper-I-Analysis/blob/fe7c5ef9163d530a4f2d4e2124a01e2048f25999/notebooks/temp_lum_comparisons/sdss_comparisons.ipynb}{Jupyter notebooks}

\subsubsection{Validation of \xga{} derived global $T_{\rm X}$ values}
\label{subsubsec:txtests}

%number returned (why some missing or added)
%comparison of best fit values
%comparison of errors

%KR: add something about the adaptability of XGA

\begin{figure*}
    \centering
    \begin{tabular}{cc}
    \includegraphics[width=1.0\columnwidth]{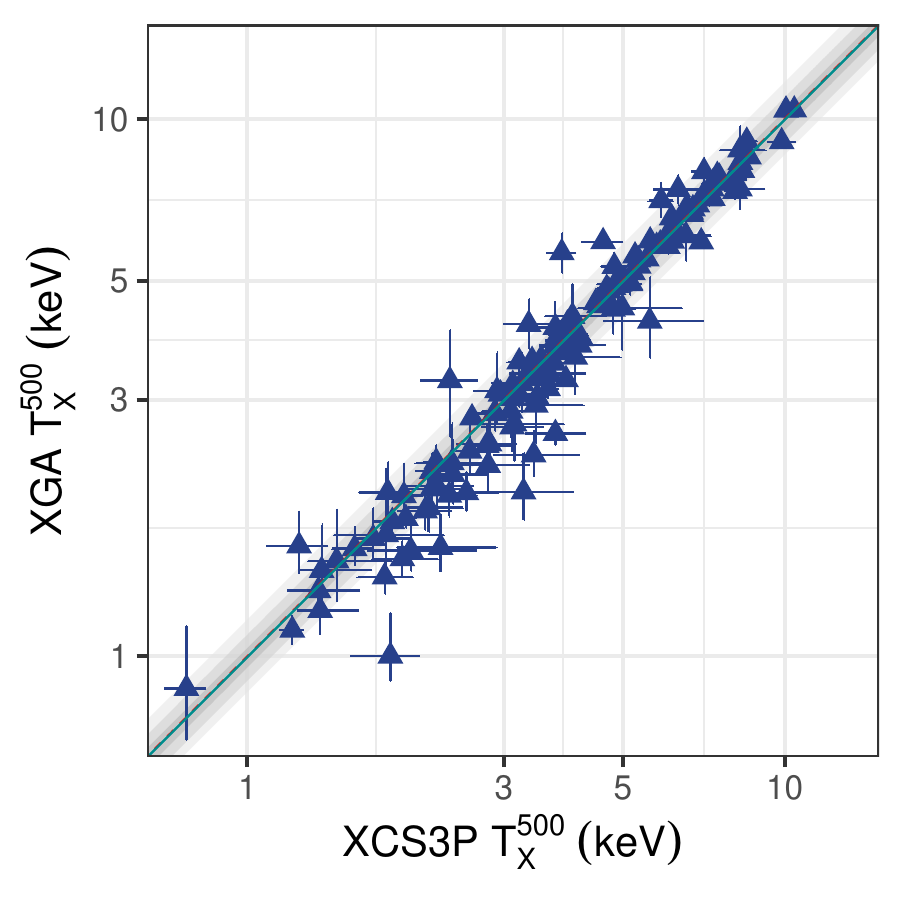} &
    \includegraphics[width=1.0\columnwidth]{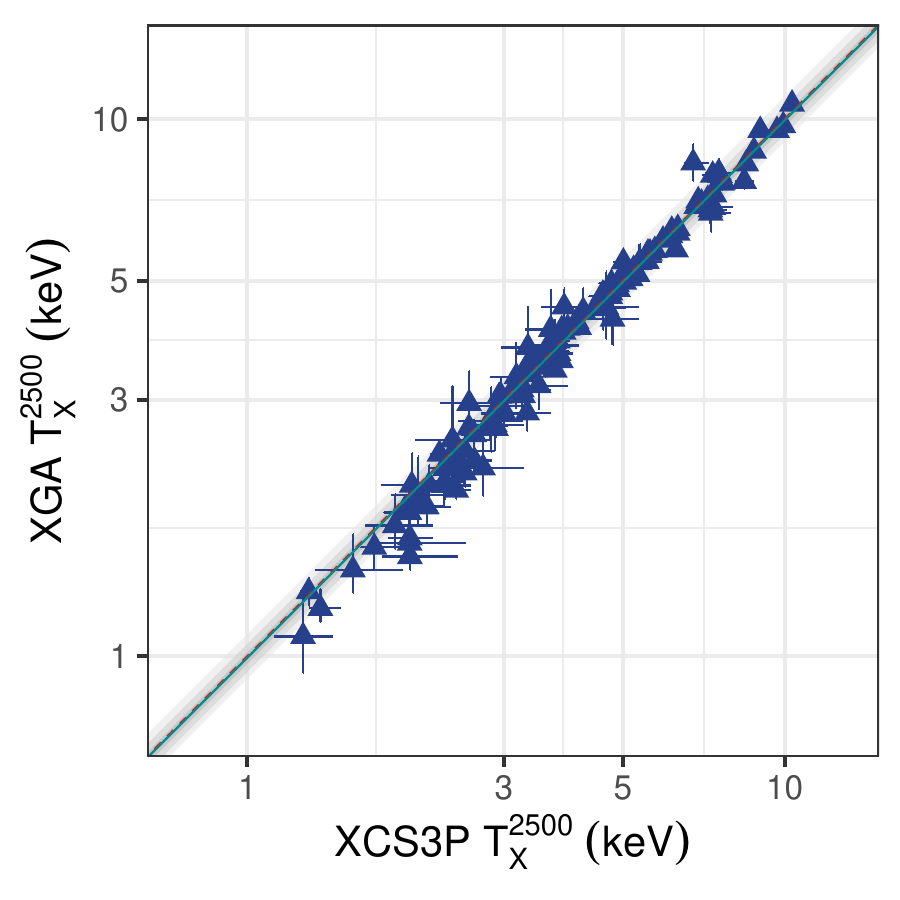} \\
    (a) & (b) \\
    \end{tabular}
    \caption[Tx SDSSRM]{The \xga{} APEC temperatures versus the \texttt{XCS3P} 
    APEC temperatures within $R_{500}$ (a) and $R_{2500}$ (b) for the SDSSRM-XCS sample \citep[as taken from][]{xcsgiles}.  The red dashed line represents the 1:1 relation (note however, the dashed line may be hidden as the best-fit line is overplotted).}
    \label{fig:xcstempcomp}
\end{figure*}

\textbf{SDSSRM-XCS:}
%\label{subsubsec:xcs3ptx}
The temperature comparison is shown in Figure~\ref{fig:xcstempcomp} for both the $R_{2500}$ (a) and $R_{500}$ (b) apertures.  The best-fit normalisations are 0.99$\pm$0.00 and 1.00$\pm$0.01 for $T_{\rm X}^{500}$ and $T_{\rm X}^{2500}$ respectively.  This highlights that the \xga{} and G22 values are in excellent agreement. This is expected, given the similarities in the input data and the methodology.
There is some small amount of scatter around the 1:1 relation, which can be attributed to the small differences in the method (see Section~\ref{subsubsec:SDSSRMsim}). 

\begin{figure}
    \centering
    \includegraphics[width=1.0\columnwidth]{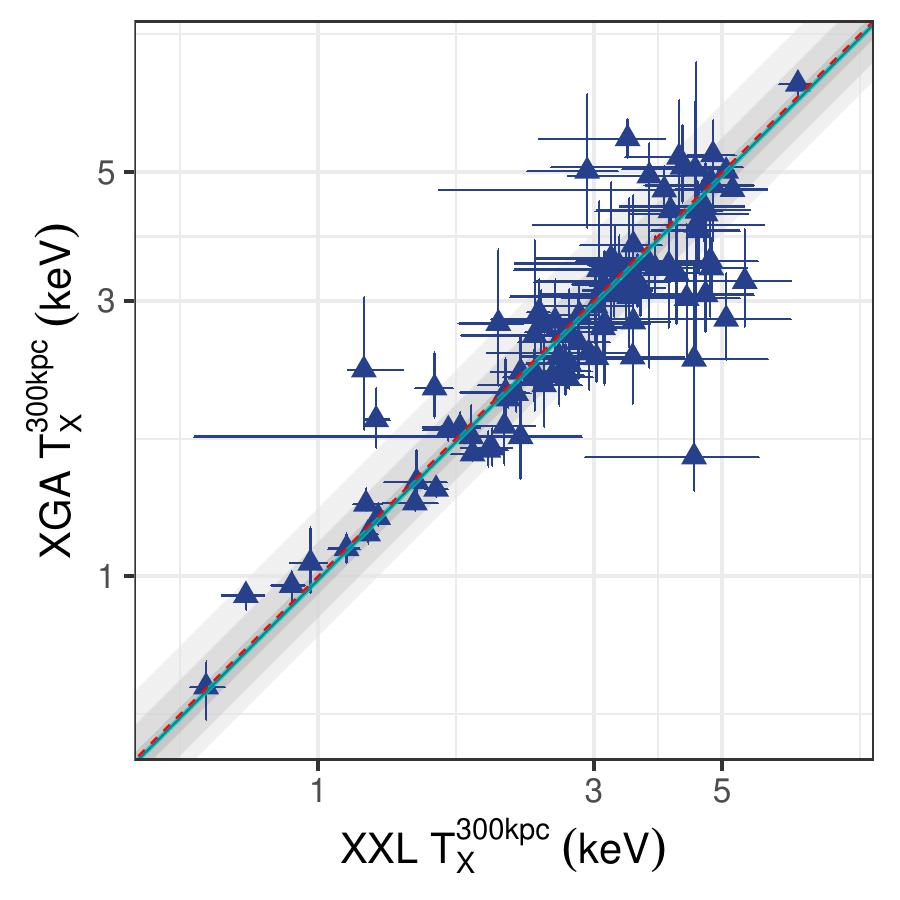}
    \caption[T comp XXL]{A comparison of global X-ray temperature values measured by \xga{} and XXL analyses \citep[as taken from][]{xxllt}, for the XXL-100-GC sample (see Section~\ref{subsubsec:txtests} for details).  The 1:1 relation is highlighted by the red dashed line.
    %comment: task: remove the Lx figure
    % b) shows the equivalent for the 0.5-2.0keV luminosity measurements.
    }
    \label{fig:xxltxcomp}
\end{figure}

\noindent \textbf{XXL-100-GC:}
%\label{subsubsec:xxltx}
The \xga{} analysis yielded T$_{\rm X}^{\rm 300kpc}$ values for \XXLTxvalues{} of the 99 clusters in the input sample. The temperature comparison is shown in Figure~\ref{fig:xxltxcomp}. The best-fit normalisation of 0.99$\pm$0.01 shows that the \xga{} and \cite{xxllt} values are in excellent agreement, despite the large number of differences between the two approaches (see Section~\ref{subsubsec:XXLsim}).

\begin{figure}
    \centering
    \includegraphics[width=1.0\columnwidth]{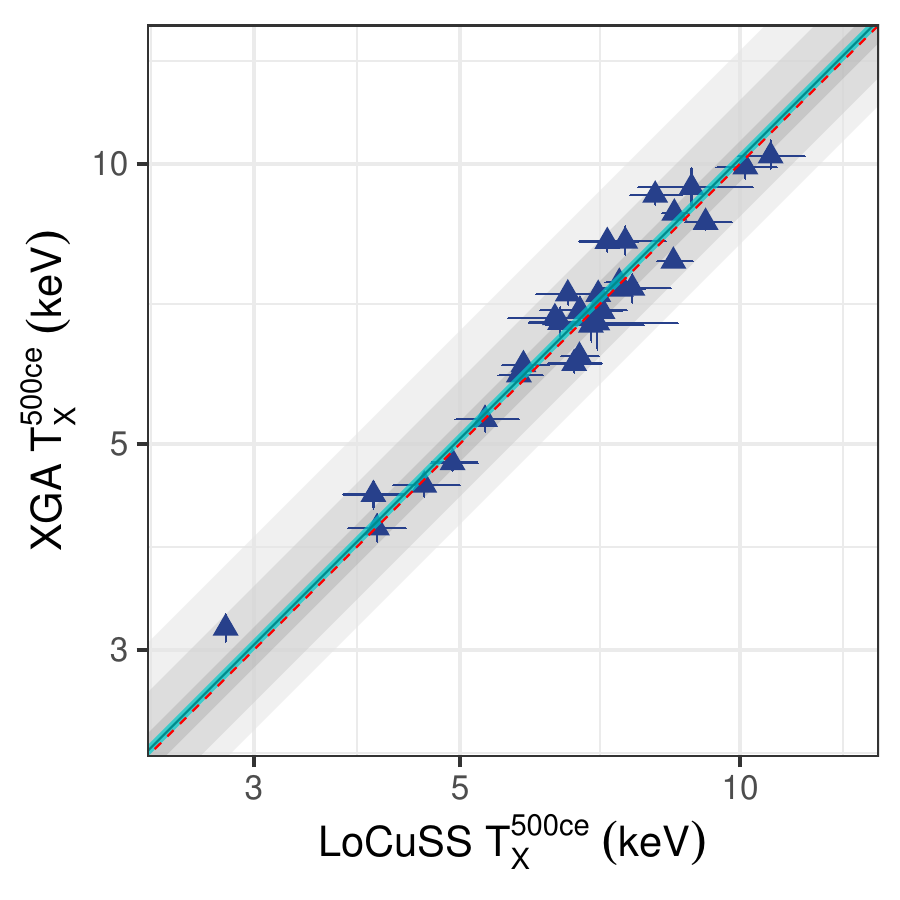}
    \caption[T comp Locuss]{A comparison of global, core-excised, X-ray $R_{500}$ temperatures measured by \xga{} and LoCuSS analyses, for a subset of the LoCuSS High-$L_{\rm{X}}$ sample. The temperatures are measured within core excluded weak-lensing $R_{500}$ values (within [0.15-1]$R_{500}$), presented in \cite{locusstemps}.}
    \label{fig:locusstxcomp}
\end{figure}

\noindent \textbf{LoCuSS High-$\pmb{L_{\rm{X}}}$}:
%\label{subsubsec:locusstx}
The \xga{} analysis yielded global, core-excised, temperatures values for all 32
of the input sample with \xmm{} results in \cite{locusstemps}. The temperature comparison is shown in Figure~\ref{fig:locusstxcomp}.  The best-fit normalisation is 1.01$\pm$0.01, showing that the \xga{} and \cite{locusstemps} values are in excellent agreement, despite the large number of differences between the two approaches (see Section~\ref{subsubsec:LoCuSSsim}).

%The temperatures which we compare to for these clusters are core-excised (measured within 0.15-1R$_{500}$), 

% The cluster spectra are centered on RASS positions reported in \cite{locusshydro}.

\subsubsection{Validation of \xga{} derived $M_{\rm gas}$ values}
\label{subsec:compgasmasses}

\begin{figure}
    \centering
    \includegraphics[width=1.0\columnwidth]{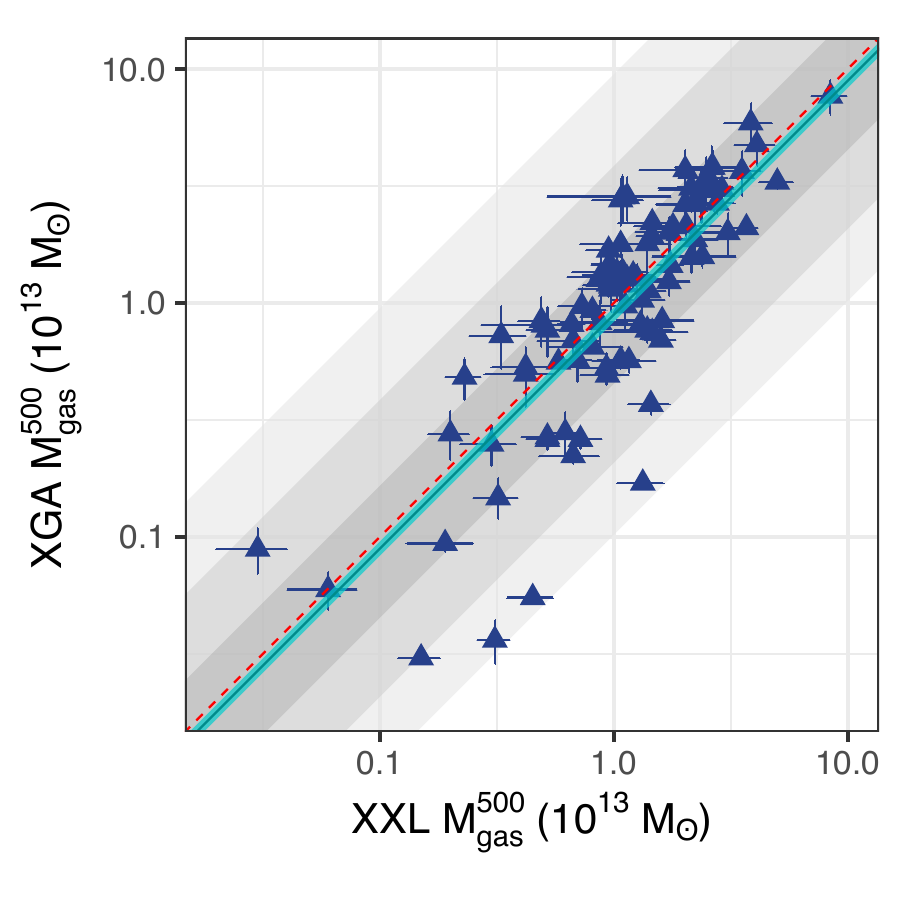}
    \caption[mgas comparison XXL]{A one-to-one comparison of gas masses measured for the XXL-100-GC cluster sample, within $R_{500}$, by \cite{xxlbaryon} and an \xga{} reanalysis. $S_B$ profiles were fitted with Beta models, density profiles with King models. Contains measurements for 91 of 96 XXL-100-GC clusters analysed with \xga{}.}
    \label{fig:compxxlgm}
\end{figure}

\noindent\textbf{XXL-100-GC:}
The \xga{} analysis yielded M$_{\rm gas}^{500}$ values for \XXLgasmassvalues{} (of the 99) clusters contained in the XXL-100-GC sample. 
%Each cluster's neutral hydrogen column density is used in the conversion factor calculation, as is its redshift, and we assume a metallicity of 0.3~$Z_{\odot}$. 
Figure~\ref{fig:compxxlgm} shows the comparison between \xga{} measured and gas mass estimates published in \cite{xxlbaryon}.  While the best-fit normalisation of 0.89$\pm$0.06 highlights the \xga{} values are $\sim$10\% lower than those measured by \cite{xxlbaryon}, the difference is $<$2$\sigma$. Therefore, there is broad agreement between the two analyses. This is encouraging considering the significant differences in density measurement between the two samples (see Section~\ref{subsubsec:XXLsim}). 

\begin{figure*}
    \centering
    \begin{tabular}{cc}
    \includegraphics[width=1.0\columnwidth]{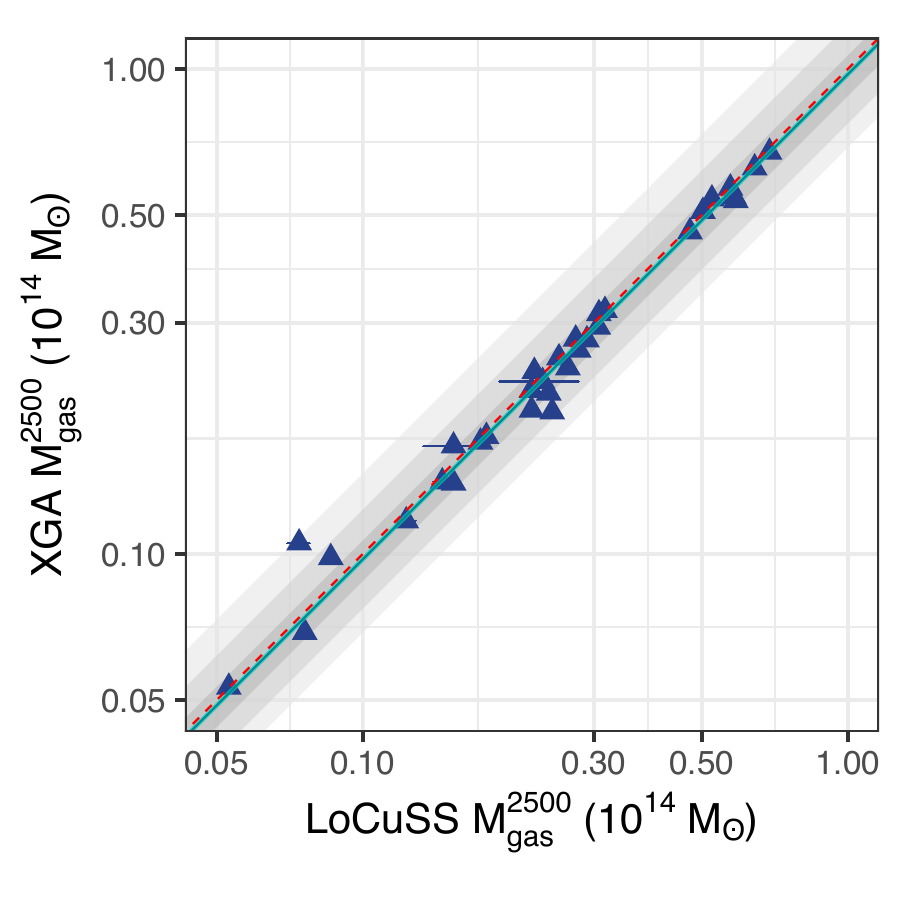} &
    \includegraphics[width=1.0\columnwidth]{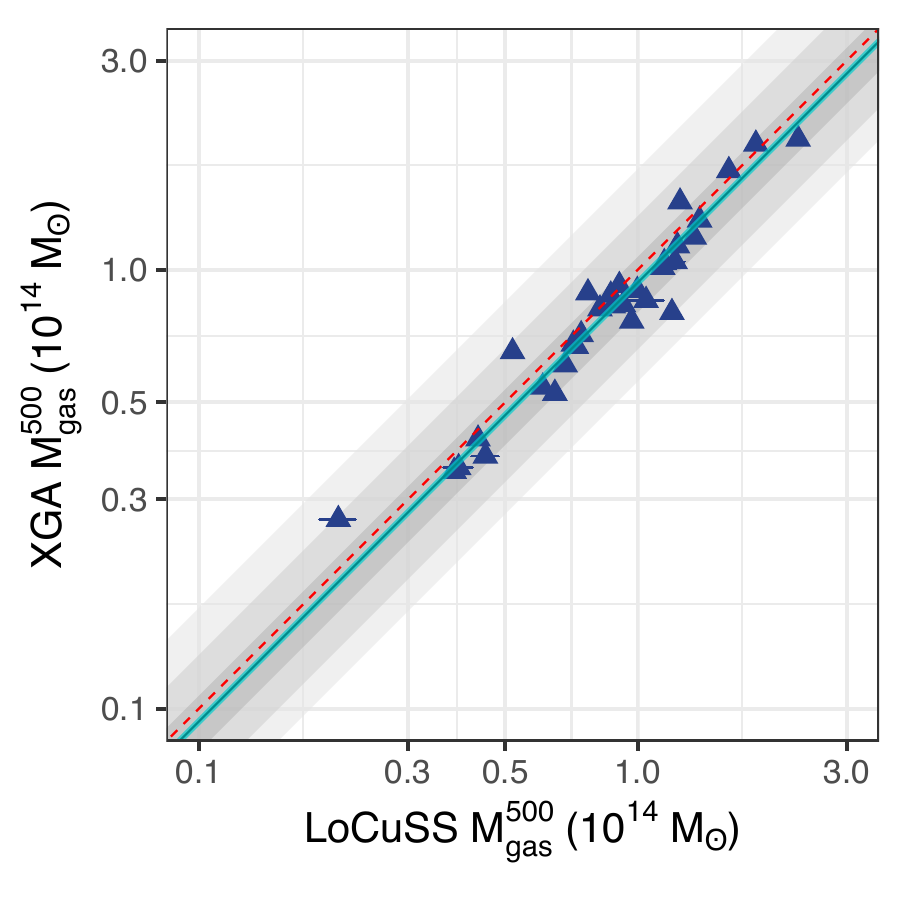} \\
    (a) & (b) \\
    \end{tabular}
    \caption[mgas comparison, locuss]{A one-to-one comparison of gas masses measured for the LoCuSS High-$L_{\rm{X}}$ cluster sample, within LoCuSS measured $R_{2500}$ (a) and $R_{500}$ (b) values, by \cite{locusshydro} and an \xga{} reanalysis.}
    \label{fig:complocussgm}
\end{figure*}

\noindent \textbf{LoCuSS High-$\pmb{L_{\rm{X}}}$}:
The \xga{} analysis yielded M$_{\rm gas}^{500}$ values for \LOCUSSgasmassvaluesxga{} (of \LOCUSSgasmassvaluesoriginal{}) LoCuSS High-$L_{\rm{X}}$ clusters.  Note that the one cluster missing (RXCJ1212) Comparisons of the gas masses calculated within R$_{2500}$ ($M_{\rm gas}^{2500}$) and R$_{500}$ ($M_{\rm gas}^{500}$) are presented in Figure~\ref{fig:complocussgm} (a) and (b) respectively.  The normalisations of the fits are 0.97$\pm$0.02 and 0.93$\pm$0.02 for $M_{\rm gas}^{2500}$ and $M_{\rm gas}^{500}$ respectively.  This highlights that the $M_{\rm gas}^{2500}$ values are consistent, however, the XGA measured $M_{\rm gas}^{500}$ values are on average 7\% lower than the LoCuSS values (significant at the 3.5$\sigma$ level).  

\subsubsection{Validation of \xga{} derived $M_{\rm hydro}$ values}
\label{subsec:mhydrocomp}

\begin{figure*}
    \centering
    \begin{tabular}{cc}
    \includegraphics[width=1.0\columnwidth]{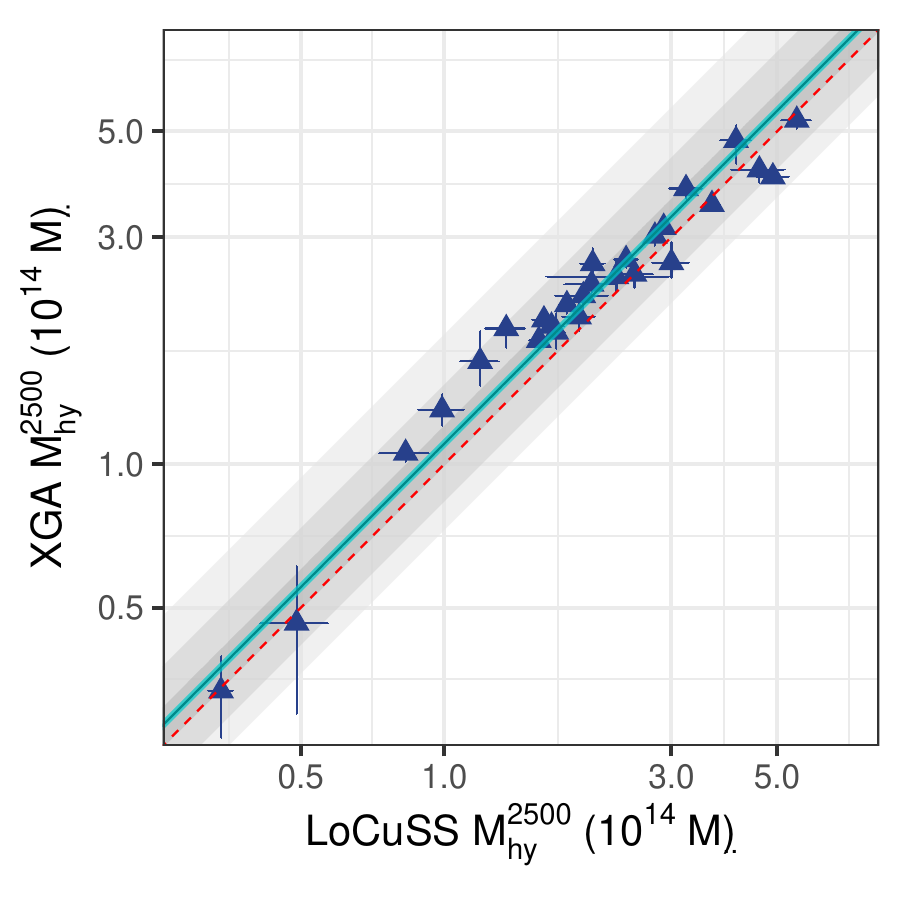} &
    \includegraphics[width=1.0\columnwidth]{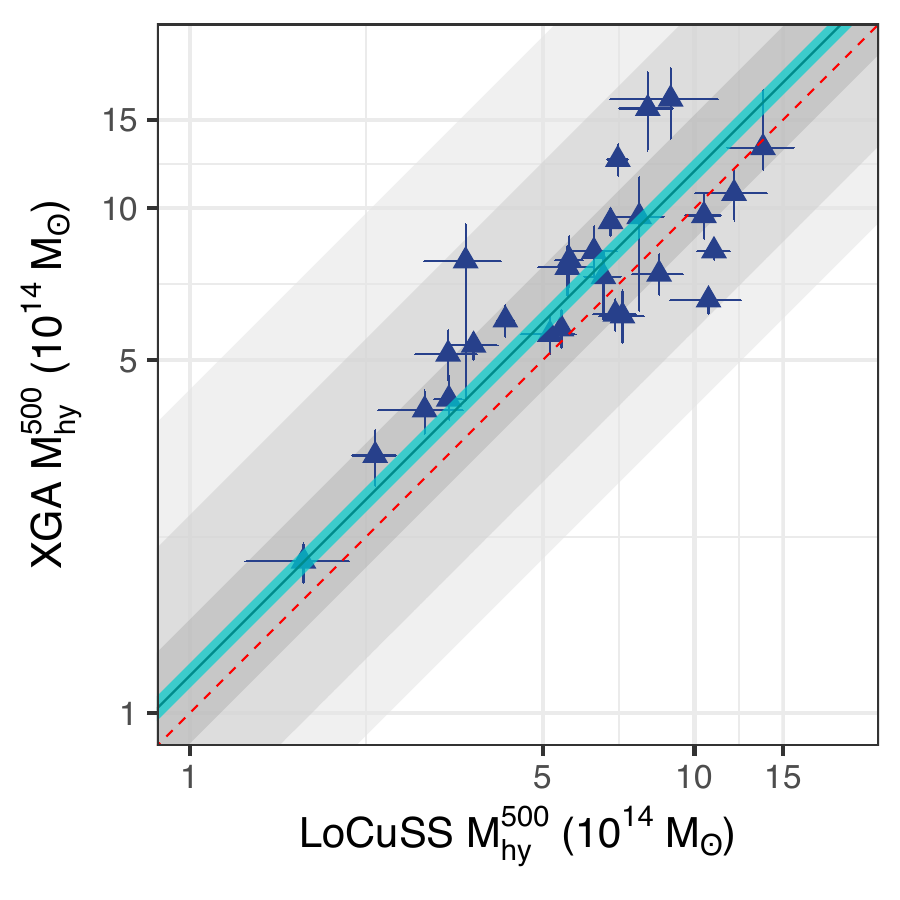} \\
    (a) & (b) \\
    \end{tabular}
    \caption[hmass comparison]{A one-to-one comparison of hydrostatic masses measured for the LoCuSS High-$L_{\rm{X}}$ cluster sample, within LoCuSS measured $R_{2500}$ (a) and $R_{500}$ (b) values, by \cite{locusshydro} and an \xga{} reanalysis.}
    \label{fig:complocusshym}
\end{figure*}

\textbf{LoCuSS High-$\pmb{L_{\rm{X}}}$}:
Figure~\ref{fig:complocusshym} shows a comparison between the \xga{} and LoCuSS hydrostatic masses for $M_{\rm{hy}}^{2500}$ (a) and $M_{\rm{hy}}^{500}$ (b). We successfully measure masses for \LOCUSShydromassvaluesxga{} galaxy clusters (of \LOCUSSTxvaluesorginal{}) in the LoCuSS High-$L_{\rm{X}}$ sample, \LOCUSShydromassvaluesoriginalxgaoverlap{} of which have \xmm{} hydrostatic masses measured by LoCuSS (which are compared in Fig.~\ref{fig:complocusshym}). When compared to the LoCuSS masses (specifically those measured by \xmm{}), we find \xga{} measured masses 10\% and 19\% higher than LoCuSS for $M_{\rm{hy}}^{2500}$ and $M_{\rm{hy}}^{500}$ respectively.  While the difference is significant at the 3.3$\sigma$ level for $M_{\rm{hy}}^{2500}$, the difference is $<$3$\sigma$ for $M_{\rm{hy}}^{500}$.

\subsection{Discussion}
%(\S\ref{subsubsec:xcs3ptx}, \ref{subsubsec:xxltx}, \ref{subsubsec:locusstx})
From the tests presented above, we can conclude that the \xga{} (almost\footnote{The data checks described in Section~\ref{subsec:manualdetreg} require human intervention.}) automated batch approaches to extracting estimates of $T_{\rm X}$, $M_{\rm gas}$ and $M_{\rm hydro}$ are robust (see Figures~\ref{fig:xcstempcomp}-\ref{fig:complocusshym} and Table \ref{tab:compfit}). In only 2 (out of 9) of the tests was the normalisation more than $3\sigma$ from unity. We have also effectively demonstrated the flexibility of our tools by matching as closely as possible the original analysis methodologies of three different samples; our use of the LoCuSS High-\lx{} and XXL-100-GC samples in particular also show that \xga{} can adapt to high and low signal-to-noise observations. All this illustrates that \xga{} can be used to easily derive properties for large sets of galaxy clusters. Uniquely, we have made \xga{} open and available for entire community to use, and we also make every effort to encourage its use by supplying extensive documentation and examples.

Where there are offsets, these are likely explained by the differing approaches taken during the analyses. In future work, we will further optimise the \xga{} method using mock observations of clusters extracted from hydrodynamic simulations \citep[such as e.g.,][]{300sims,milleniumTNGclusters}, i.e. where the ``true'' gas and halo masses are known. This will also allow us to estimate the systematic errors introduced from simplifying assumptions, such as spherical symmetry and hydrostatic equilibrium. We also note that we have derived temperature profiles for all galaxy clusters for which a hydrostatic mass has been measured, whereas some previous work \citep[e.g.][]{locusshydro} assumes a temperature profile in cases where there are low total counts. As we apply \xga{} to the measurement of hydrostatic masses in the future we will work to implement improvements to our method, including modelling of the X-ray background to better support the analysis of low-redshift systems, and more sophisticated approaches to the deprojection of temperature profiles.

\section{Mass analysis of the SDSSRM-XCS sample}
%comment:  It would be ironic, but have we inadvertently introduced some miscentering effect of our own by not re-measuring richness around the X-ray centre
\label{sec:results}
Here we present hydrostatic masses for a subset of galaxy clusters in the SDSSRM-XCS sample presented by \cite{xcsgiles}.  

\subsection{Masses for a subset of the SDSSRM-XCS sample}
\label{subsec:newmasses}

We attempt to measure masses for all 150 galaxy clusters in the SDSSRM-XCS sample.  Measurements presented in this section are performed using \xga{}, within the $R_{500}$ and $R_{2500}$ values measured by G22. In Section~\ref{subsubsec:txtests} we demonstrated that the global \xga{} \tx{} measurements are consistent with the original G22 analysis when measured within identical apertures (see Fig.~\ref{fig:xcstempcomp}). 

%We do not expect to be able to measure masses for all of these clusters, as measuring temperature profiles requires significant X-ray counts, but we shall attempt measurements using \xga{}'s automated routines. This will demonstrate its robustness and ability to deal with failed measurements.

Temperature and density profiles were generated out to $R_{500}$ (from G22), with the hydrostatic masses estimated from these profiles at $R_{500}$ and $R_{2500}$. Density profiles were generated as described in Section~\ref{subsec:densityProf}, and temperature profiles were generated and selected as described in Section~\ref{subsec:gastempprof} from annular spectra (described in Section~\ref{subsec:xgaspecs}).

We are able to successfully measure hydrostatic masses for \sdssrmmasses{} of the 150 galaxy clusters in the SDSSRM-XCS sample. The temperature profiles used for these measurements have between 4 (the minimum number allowed, see Section~\ref{subsec:xgaspecs}) 
and 26 annuli. This set of hydrostatic masses adds significant value to the SDSSRM-XCS sample, as well as representing one of the largest samples of cluster hydrostatic masses available. As each mass is measured individually, rather than derived through the stacking of different clusters (as is usually the case for weak lensing cluster masses), there are many possible uses for the sample. Also, as SDSSRM-XCS clusters were selected from optical/NIR via the SDSS redMaPPer catalogue \citep[][]{redmappersdss}, this sample allows for powerful multi-wavelength studies of how cluster mass is related to optical properties.

\subsection{Comparison of SDSSRM-XCS masses with the literature}
\label{subsec:sdssrm_hydro_comp}

\begin{figure*}
    \centering
    \begin{tabular}{cc}
    \includegraphics[width=1.0\columnwidth]{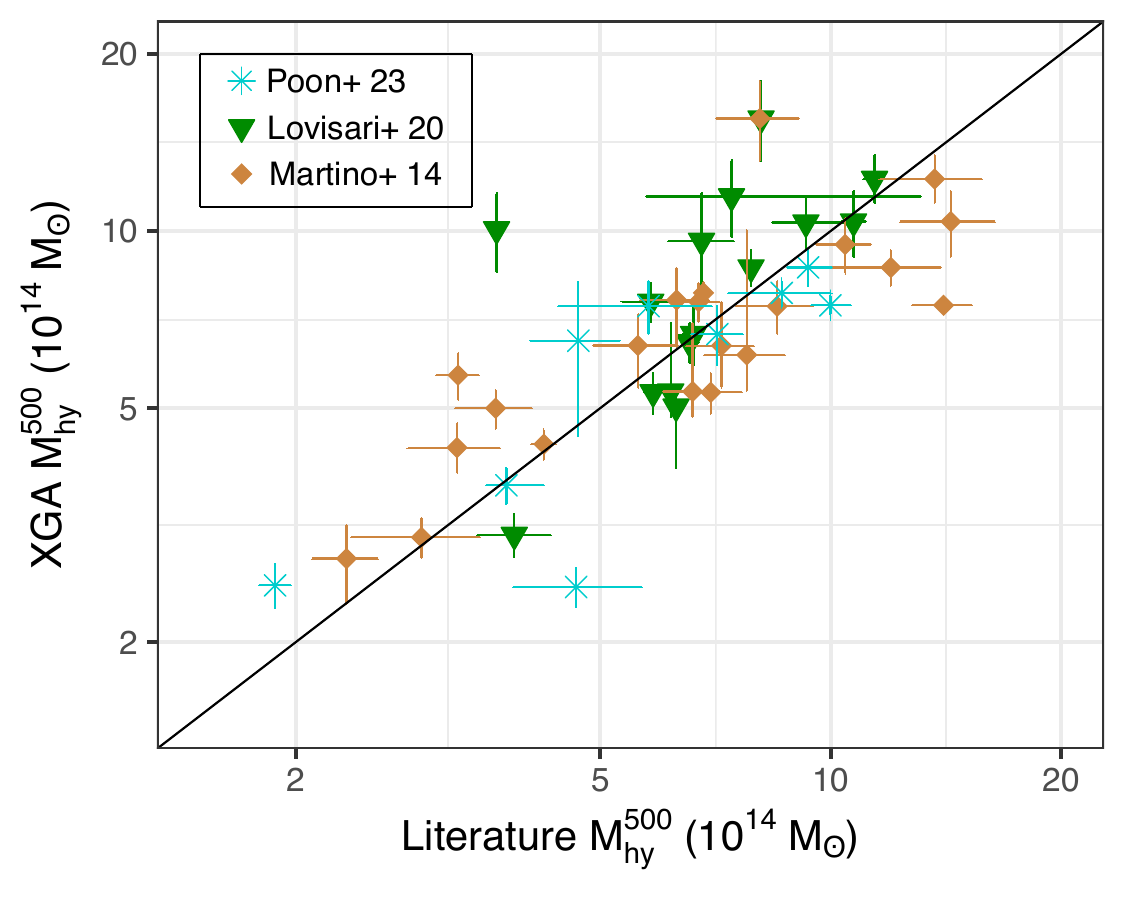} &
    \includegraphics[width=1.0\columnwidth]{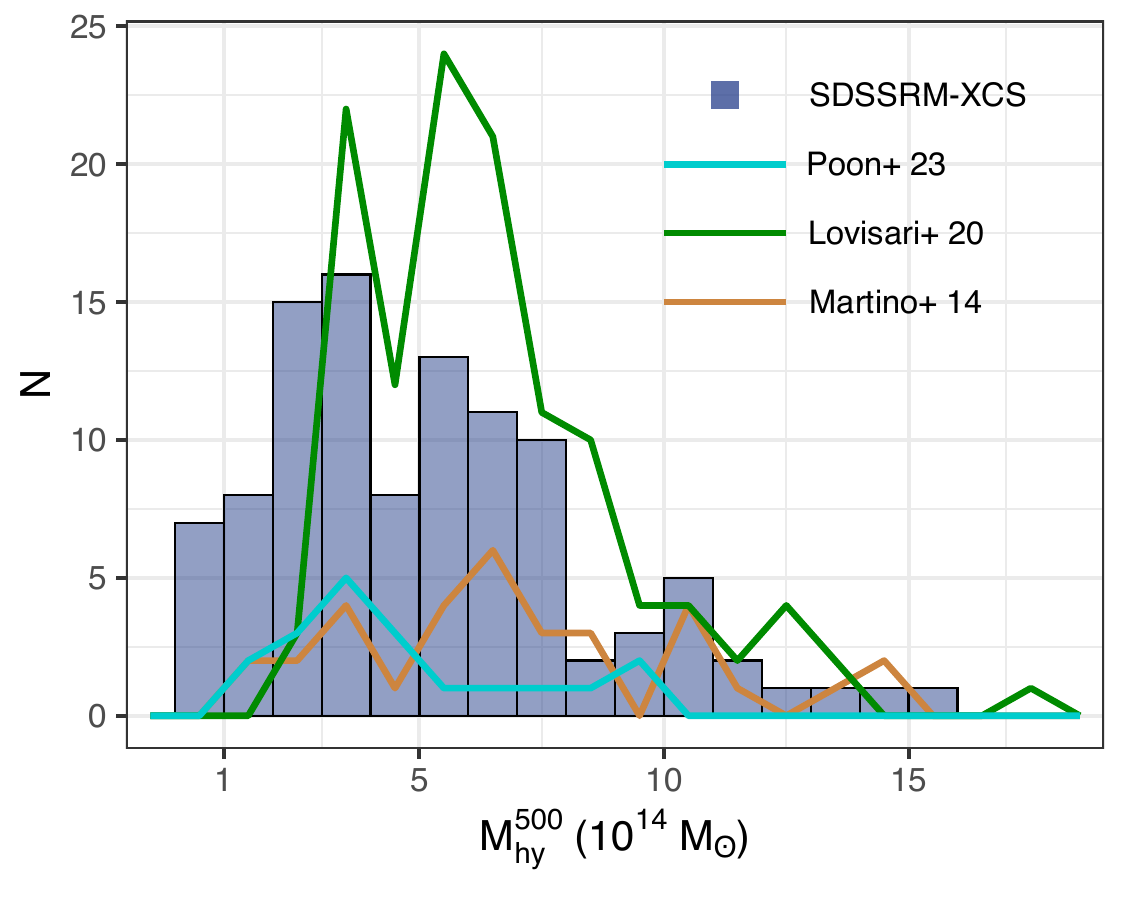} \\
    (a) & (b) \\
    \end{tabular}
    \caption[mass comparison]{(a) A one-to-one comparison of hydrostatic masses measured for the SDSSRM-XCS cluster sample using \xga{}, compared to \citet[][brown diamonds]{locusshydro}, \citet[][green triangles]{lovisarimasses} and \citet[][cyan crosses]{2023MNRAS.520.6001P}. (b) Mass distribution of the 102 SDSSRM-XCS clusters with M$^{500}_{\rm hy}$ values (lightblue histogram).  Mass distributions from \citet[][brown histogram, using only the {\em XMM} based masses]{locusshydro}, \citet[][green]{lovisarimasses} and \citet[][cyan]{2023MNRAS.520.6001P} are given for comparison.}
    \label{fig:sdssrm_mass_comp}
\end{figure*}

Here we compare the masses estimated for the SDSSRM-XCS sample in Section~\ref{subsec:newmasses} to values taken from the literature.  We compare the M$^{500}_{\rm hy}$ values to those given in \cite{locusshydro}, \cite{lovisarimasses} and \cite{2023MNRAS.520.6001P}.  The \cite{locusshydro} masses are from the LoCuSS sample as discussed in Section~\ref{subsec:samples}.  The \cite{lovisarimasses} sample contains 120 {\em XMM} derived hydrostatic masses from {\em Planck} detected clusters (spanning a redshift range of $0.059 < z < 0.546$).  Finally, \cite{2023MNRAS.520.6001P} estimate hydrostatic masses using {\em XMM} for 19 clusters contained in the Meta-Catalog of X-Ray Detected Clusters of Galaxies \citep[][]{mcxc_sample} within the Hyper Suprime-Cam Subaru Strategic Programme field \citep[][]{HSC_SSP}.  The \sdssrmmasses{} SDSSRM-XCS clusters with M$^{500}_{\rm hy}$ values are matched to the samples of clusters used in the works listed, resulting in 21, 16 and 9 clusters matched to \cite{locusshydro}, \cite{lovisarimasses} and \cite{2023MNRAS.520.6001P} respectively.  The comparison of the masses is plotted in Figure~\ref{fig:sdssrm_mass_comp}(a).  The black line represents the 1:1 relation, highlighting a broad consistency of the masses measured in this work and those reported in the literature.

In Figure~\ref{fig:sdssrm_mass_comp}(b), we show the number of SDSSRM-XCS clusters per M$^{500}_{\rm hy}$ bin (given by the blue histogram).  The M$^{500}_{\rm hy}$ distributions from \cite{locusshydro} and \cite{lovisarimasses} are given by the brown and green spline curves respectively.  While the SDSSRM-XCS sample clearly contains more masses than the \cite{locusshydro} sample, clusters in \cite{locusshydro} are selected by imposing a luminosity limit on clusters detected in the RASS, reducing the number of available clusters for study.  The \cite{lovisarimasses} sample contains a larger number of masses overall, the SDSSRM-XCS sample provides many more low-mass cluster estimates.  However, we note that this is due to the {\em Planck} selection of the clusters in \cite{lovisarimasses}, which will select higher mass clusters.  Additionally, \cite{lovisarimasses} provide masses for 17 clusters for which their masses were estimated using $<$4 temperature profile bins.  For the analysis of the SDSSXCS-RM sample, we required a minimum of 4 bins in the temperature profile for a mass determination.

\section{Summary and next steps}
\label{sec:conclusions}
Finally, in this section we present summaries of the work in this paper, discussions of the implications and results, and goals for related future work.

\subsection{Summary}
\label{subsec:discussion}

This work introduces an automated method, and tool (\xga{}), for measuring hydrostatic masses (and many other X-ray properties) of large samples of galaxy clusters. We summarise this tool, and the approaches taken to measure the various properties that are required to determine hydrostatic masses, in the first part of the paper. Two demonstration clusters (\clustone{} and \clusttwo{}) are used to help illustrate the different steps and data products required. This includes the measurement of density and 3D temperature profiles. As part of this we showcase and explain \xga{}'s PSF correction feature, which takes into account the spatially varying nature of \xmm{} PSFs, and is relevant to the measurement of density profiles.

We then make use of three samples of galaxy clusters, SDSSRM-XCS, XXL-100-GC, and LoCuSS High-$L_{\rm{X}}$ to explore the efficacy of measurements of galaxy cluster properties produced by our new software tool, \xga{}. Once we have demonstrated that \xga{} is reliable, we find and present new measurements of hydrostatic mass for the SDSSRM-XCS sample. In summary:

\begin{itemize}[leftmargin=*, labelindent=2pt, itemsep=1ex]
  \item[--] Comparisons of \tx{} measurements of the SDSSRM-XCS sample from an existing XCS analysis \citep[][]{xcsgiles} to those measured by \xga{} demonstrate close agreement. This holds true for measurements within both $R_{500}$ and $R_{2500}$, with $R_{2500}$ measurements appearing to show less scatter. Close agreement of these measurements is expected because they use the same event lists and region files, as well as very similar analysis techniques. We quantify the comparisons by plotting \xga{} results against literature value and fitting a fixed-slope power law; all normalisations are consistent with unity.
  
  \item[--] Similar comparisons of \tx{} to literature values measured for the XXL-100-GC \citep{xxllt} and LoCuSS High-$L_{\rm{X}}$ \citep{locusstemps} samples showed good agreement, and also demonstrated \xga{}'s ability to perform measurements with different energy limits, with metallicity left free to vary, and different cosmologies. Model fits again showed that the comparison normalisations are consistent with unity.
 
%  \item[--] This work has highlighted that small analysis choices (such as the energy limits used for spectral fits) can have significant effects on global measurements made for a sample of galaxy clusters. This indicates that samples from literature should not be naively combined, but rather that a reanalysis should be performed with consistent analysis choices; \xga{} makes that a realistic possibility. 

  \item[--] We then began to test values derived from radial profiles, starting with gas masses. Comparisons of gas mass values measured for the XXL-100-GC \citep{xxlbaryon} and LoCuSS High-$L_{\rm{X}}$ \citep{locusshydro} samples, and reanalyses using \xga{}, demonstrated that \xga{} produces gas mass measurements consistent with past work. The XXL comparison indicated an small ($\sim$10\%) offset between the two analysis, however, the difference was not significant. The LoCuSS comparison demonstrated a 1:1 relation between gas masses measured within LoCuSS $R_{2500}$ values, but that $R_{500}$ gas masses are slightly systematically larger when measured by \xga{} versus the original work. 
  
  \item[--] Our final comparison makes use of the hydrostatic masses measured within $R_{2500}$ and $R_{500}$ for the LoCuSS High-$L_{\rm{X}}$ sample by \cite{locusshydro}. The $M^{\rm{hy}}_{2500}$ comparison lies close to the 1:1 line, with minimal scatter. The $M^{\rm{hy}}_{500}$ comparison is also broadly compatible with the 1:1 line, though the scatter and uncertainties are increased when compared to the $M^{\rm{hy}}_{2500}$ plot. From this comparison, and the others presented in this work, we conclude that \xga{} mass measurements are consistent with previous work and can be relied on.
  
  \item[--] Finally, we present new measurements of hydrostatic mass for the SDSSRM-XCS galaxy cluster sample. In total, we measure hydrostatic masses for \sdssrmmasses{} (out of 150) clusters, which is comparable to the largest consistently analysed and measured X-ray based hydrostatic masses in the literature. 

  \item[--] We have demonstrated that \xga{} is a mature tool that is capable of measuring many key properties of galaxy clusters, and easily replicating existing analyses, producing comparable results. Any \xga{} analysis is easy to scale to any size of sample. It also allows colleagues without specialist X-ray experience to carry out their own cluster mass analyses.

    %comment: worth saying in Discussion? KR "The region adjustment process took less than four hours."
  
\end{itemize}

\subsection{Future Work}
\label{subsec:future}

Here we detail some of the work that we have planned for our new implementation of hydrostatic mass measurement, as well as for the samples of masses we have (and will) measure.

\subsubsection{Mass-Observable Relations}
\label{subsubsec:massobsrel}

Whilst we have demonstrated the utility of \xga{} for measuring hydrostatic masses for large samples of galaxy clusters, and measured masses for a significant number of clusters using archival data, we have not yet constructed mass-observable relations. As we discussed in the introduction, such relations are an invaluable independent measure of the normalisation, slope, and scatter of cluster masses with other observables. The next paper in this series will focus on constructing such relations for a larger set of galaxy clusters, and for both X-ray and optically derived observables; they will be extremely useful for the next-generation of cluster cosmology being enabled by new telescopes and missions.

\subsubsection{Other Samples}
\label{subsubsec:othersamples}

We will produce mass measurements and scaling relations for other samples of galaxy clusters, selected from optical/NIR surveys such as DES, and from the ACT-DR5 cluster catalogue. This shall not only add a significant number of cluster masses to our sample, but will also open the possibility of constructing scaling relations between X-ray mass and Sunyaev–Zeldovich properties.

\subsubsection{Temperature Profile Methodology}
\label{subsubsec:3dtempmethimprov}

In this initial work we have used a relatively simple methodology to `deproject' the measured temperature profiles and infer the three-dimensional radial temperature structure of the galaxy clusters. More sophisticated techniques could be applied to this problem \citep[such as deep learning deprojection,][]{chexmate_deprojection}, which might improve the temperature profiles measured by \xga{}, and thus the hydrostatic masses. This will be achievable through the existing framework of \xga{}, and future work with different techniques to measure 3D temperature profiles would also provide a systematic comparison between the measured profiles, and inferred masses.

We will also improve the sophistication of the spectral fits used to generate the projected temperature profiles, taking more care to account for the effects of the \xmm{} PSF on the photons detected within each annulus.

\subsubsection{Multi-mission X-ray Analyses}
\label{subsubsec:multimission}

Planned additions to the \xga{} software package will enable support for X-ray telescopes other than \xmm{}; e.g. \chandra, \erosita, and {\em XRISM}. This will allow a user to draw on multiple archives of X-ray observations for analysis of samples, increasing the likelihood that samples selected from other wavelengths will have a serendipitous X-ray observation.

Multi-mission support will also allow us to design joint analyses that take advantage of the unique capabilities of each telescope. Joint
analyses with all available X-ray data should be-
come routine, rather than requiring special effort,
and \xga{} can provide that capability. 

We will also include support for simulated telescopes to enable preparatory work for future missions; e.g. {\em Athena}, {\em Lynx}, and {\em AXIS}. As such we will produce catalogues of hydrostatic masses measured with multiple current telescopes (\xmm{} and \chandra{}), as well as explore how we may exploit the capabilities of planned telescopes. Work on simulated clusters will also include exploration of the hydrostatic mass bias from true mass.

\subsubsection{Accounting for non-thermal pressure support}
\label{subsubsec:ntp}

Hydrostatic masses are biased from true masses due to assumptions made during the derivation of Equation~\ref{eq:hydromass}. The assumption of hydrostatic equilibrium implies that all pressure support is provided by the thermal gradient of the ICM, which is not the case. Various processes also help to balance gravitational collapse, and together are often referred to as non-thermal pressure (NTP) support. Methods to measure a total mass from X-ray observations exist \citep[e.g.][]{xcop_ntp}, and we aim to apply them to large samples of clusters such as SDSSRM-XCS. By taking NTP support into account hydrostatic masses can become even more competitive with other direct mass measurement methods (such as individual weak lensing).

\section*{Acknowledgements}
We made use of TOPCAT \citep[][]{topcat} in various parts of this project. The X-ray analysis module developed by XCS (\xga{}) makes significant use of Astropy \citep[][]{astropy1, astropy2}, NumPy \citep[][]{numpy}, Matplotlib \citep[][]{matplotlib}, and pandas \citep[][]{pandassoftware,pandaspaper}. \xga{} also uses GetDist \citep[][]{getdist} to produce corner plots.

DT, KR, and PG acknowledge support from the UK Science and Technology Facilities Council via grants ST/P006760/1 (DT), ST/P000525/1, ST/T000473/1, and ST/X001040/1 (PG, KR). DT is also grateful for support from the National Aeronautic and Space Administration Astrophysics Data Analysis Program (NASA-80NSSC22K0476). PTPV was supported by Fundação para a Ciência e a Tecnologia (FCT) through research grants UIDB/04434/2020 and UIDP/04434/2020.

We would like to thank Judith H. Croston for her useful comments on this work, Graham Smith for providing LoCuSS data, Aswin P. Vijayan and Lucas Porth for useful discussions during the course of this research.

%%%%%%%%%%%%%%%%%%%%%%%%%%%%%%%%%%%%%%%%%%%%%%%%%%
\section*{Data Availability}
The \xmm{} data underlying this article were accessed from the {\em XMM} science archive. The derived data underlying this article are available in the article and in its online supplementary material. All analysis code is available in the accompanying GitHub repositories.

%%%%%%%%%%%%%%%%%%%% REFERENCES %%%%%%%%%%%%%%%%%%

% The best way to enter references is to use BibTeX:

\bibliographystyle{mnras}
\bibliography{xcs_mass_I} % if your bibtex file is called example.bib

% Alternatively you could enter them by hand, like this:
% This method is tedious and prone to error if you have lots of references
%\begin{thebibliography}{99}
%\bibitem[\protect\citeauthoryear{Author}{2012}]{Author2012}
%Author A.~N., 2013, Journal of Improbable Astronomy, 1, 1
%\bibitem[\protect\citeauthoryear{Others}{2013}]{Others2013}
%Others S., 2012, Journal of Interesting Stuff, 17, 198
%\end{thebibliography}

%%%%%%%%%%%%%%%%%%%%%%%%%%%%%%%%%%%%%%%%%%%%%%%%%%

%%%%%%%%%%%%%%%%% APPENDICES %%%%%%%%%%%%%%%%%%%%%

\appendix

\section{Fitting models to radial profiles in \xga{}}
\label{app:radmodfit}

\begin{table*}
\centering
\caption[]{{\small Summaries of the \xga{} 1D radial models used in this work (though others are implemented and available). Model names and descriptions of their use are included. The units of model parameters are given, as well as the start parameters used for initial fits. Finally details of the parameter priors used for the full MCMC fit are given, where $\mathcal{U}$[A, B] indicates a uniform distribution with limits A and B.}\label{tab:models}}
\vspace{1mm}
\addtolength{\tabcolsep}{-0.7em}
\begin{tabular}{ccccccc}
\hline
\hline
Model Name & Equation & Brief description & Parameters & Start Values & Priors & \\
\hline
\hline
\multirow{3}{1.5em}{\centering Beta} & \multirow{3}{17.5em}{\centering $S_X(R) =  S_0\left( 1 + \left( \dfrac{R}{R_c} \right)^{2}\right)^{-3\beta + 0.5}$} & Simple model for the & $S_{0}$ [$\rm{ct}\rm{s^{-1}}\:\rm{arcmin}^{-2}$] & 1 & $\mathcal{U}$[0, 3] & \\ 
& & surface brightness profile, & $R_{\rm{core}}$ [kpc] & 100 & $\mathcal{U}$[1, 2000] & \refstepcounter{equation}\thetag{\theequation}\label{eq:beta}\\ 
& & \cite{beta}. & $\beta$ & 1 & $\mathcal{U}$[0, 3] & \\ 

\hline

\multirow{3}{1.5em}{\centering King} & \multirow{3}{17.5em}{\centering $\rho_{g}(R) = \rho_0 \left( 1+ \left( \dfrac{R}{R_c} \right)^{2}\right)^{-3\beta}$} & A very simple model for the &  $\rho_{0}$ [$\rm{M_{\odot}}\rm{Mpc^{-3}}$] & 1 & $\mathcal{U}$[1, 10000]$\times10^{12}$ & \\ 
& & density profile; an inverse-Abel & $R_{\rm{core}}$ [kpc] & 100 & $\mathcal{U}$[0, 2000] & \refstepcounter{equation}\thetag{\theequation}\label{eq:king}\\ 
& & transform of the beta profile. & $\beta$ & 1 & $\mathcal{U}$[0, 3] & \\ 

\hline

\multirow{6}{6.5em}{\centering Simplified Vikhlinin Temperature} & \multirow{6}{18.5em}{\centering $T(R) = \dfrac{T_{0}\left(\left(\frac{R}{R_{c}}\right) + \frac{T_{min}}{T_{0}} \right)\left( \frac{R^{2}}{R_{T}^{2}} + 1 \right)^{-\frac{c}{2}}}{\left(\frac{R}{R_{c}}\right)^{a_{c}} + 1}$} & & $R_{\rm{c}}$ [kpc] & 50 & $\mathcal{U}$[10, 500]\\ 
& & Simplified \cite{vikhlininmass}  & $a_{\rm{c}}$ & 1 & $\mathcal{U}$[0, 5] \\ 
& &  temperature profile as used by & $T_{\rm{min}}$ [keV] & 3 & $\mathcal{U}$[0.1, 6] & \refstepcounter{equation}\thetag{\theequation}\label{eq:simpvikhtemp}\\ 
& &  \cite{simpvikhprofile}. & $T_{0}$ [keV] & 6 & $\mathcal{U}$[0.5, 15]\\ 
& & & $R_{\rm{T}}$ [kpc] & 200 & $\mathcal{U}$[100, 500] \\ 
& & & c & 1 & $\mathcal{U}$[0, 5] & \\
\hline
\end{tabular}
\end{table*}

Various aspects of this work require models to be fitted to radial data profiles. For this work we make use of the \texttt{emcee} ensemble MCMC sampler which has been widely used in the astronomy community \citep[][]{emcee}. This sampler runs multiple interconnected MCMC chains to explore the parameter space. We choose to use a simple Gaussian likelihood function and uninformative uniform priors. The priors are dependant upon the particular profile being fitted, as well as the model choice, and further information can be found in table \ref{tab:models}.

All profile fitting performed for this work used an \texttt{emcee} sampler instance with 20 walkers, each taking 30000 steps. We chose to generate sets of start positions for the walkers by adapting a simple method suggested in the \texttt{emcee} documentation. We use the \texttt{SciPy} \citep[][]{scipy} implementation of the non-linear least squares (NLLS) fitting method (\texttt{curve\_fit}) to fit the model to the profile. The NLLS fits use starting values for model parameters as defined in Table~\ref{tab:models}. We then find the order of magnitude of each fit parameter. These are then perturbed by drawing an $N_{\rm{walker}}\times N_{\rm{par}}$ matrix of random values from a $\mathcal{N}(0, 1)$ distribution, multiplying each random value by the order of magnitude of the respective parameter, and then adding this perturbation to the original NLLS best fit value. Before starting, the sampler checks to ensure that the start positions all fall within the allowed boundaries of the priors, and if not then they are re-drawn until they do.
% On the rare occasions that the \texttt{curve\_fit} run fails, we fall back on a maximum likelihood method where we attempt to minimise the negative log likelihood to find best fit parameters, then those parameters are perturbed in the same way to generate start positions for the walkers. 

Once the sampler run is complete, we calculate integrated autocorrelation times for the model parameters, if all autocorrelation times are more than 400 times smaller than the number of steps taken by a walker, then we find the mean autocorrelation time for all the parameters, round it up to the nearest 100, and double it to find a good number of steps to remove from each walker's chains as a burn-in period. If any of the parameters had an autocorrelation time that did not fit this criteria, then we take a brute force approach and remove the first 30\% of steps in each walker's chains. We draw 10000 random points from combined parameter chains from all walkers to create the final posterior distributions. We also measure medians of the parameter posterior distributions, as well as finding the 1$\sigma$ regions.

These fits are performed using a set of radial model classes that we have implemented in \xga{}. They provide easy access to model posterior distributions, visualisations, and predictions. \xga{} models also have implementations of useful mathematical operations such as differentiation, spherical volume integration, and inverse Abel transformation of the models \citep[using PyAbel;][]{pyabelmeth,pyabelsoft}, making use of Astropy \citep[][]{astropy1, astropy2} to provide correct units for any calculated quantities. Model instances also provide information on model parameters, support custom start parameters, and also custom priors. 

\section{Companion GitHub Repository}
\label{app:companionrepo}

The work in this paper is accompanied by a GitHub repository$^{\ref{foot:gitrepo}}$, containing the samples and code required to reproduce our work, as well as tables of results and extra figures. Here we briefly summarise the contents of the repository and provide the context which connects sections of the paper to the different files. The vast majority of files containing sample information and results are `comma separated variable' files (csv), which are human-readable and can be opened in a text editor, or more practically can be opened with software such as TOPCAT \citep[][]{topcat}, or a Python module such as Pandas \citep[][]{pandaspaper}. Our companion repository contains several directories:

\subsection{Sample Files}
\label{app:sampfiles}
This part of the repository contains files of galaxy cluster properties for the samples used in this work. The exact contents of the sample files varies, but at minimum they contain a cluster name, a central coordinate, a redshift measurement, and some measurements of X-ray cluster properties that are used as validation at some point in this work;

\begin{itemize}[leftmargin=*, labelindent=2pt, itemsep=1ex]
    \item \textbf{xcs3p\_sdssrm\_clusters.csv} - A sample file containing the properties measured for the volume complete, temperature within $R_{500}$ fractional uncertainty less than 25\%, SDSSRM-XCS sample of clusters presented by \cite{xcsgiles}. This sample is introduced in Section~\ref{subsubsec:SDSSRMsim}, and is used both for validation purposes, and for the measurement of new hydrostatic masses. The file includes X-ray centroid positions, redshifts, $R_{500}$ and $R_{2500}$ values (with uncertainties), \tx{}, and \lx{} values within the two apertures. Not all clusters in this sample have $R_{2500}$ values and associated \lx{} and \tx{} values measured within them.
    \item \textbf{xxl\_gc100.csv} - A file with information relevant to the XXL-100-GC sample of galaxy clusters, introduced in Section~\ref{subsubsec:XXLsim}. The information available includes positions, redshifts, and $R_{500}$ measurements for the clusters, as well as XXL names and quality flags. Derived properties such as \tx{} and \lx{}, measured by \cite{xxllt}, and gas mass values measured by \cite{xxlbaryon}, are also included. All data, apart from $R_{500}$ uncertainties, are taken from the XXL-100-GC VizieR table; the radius uncertainties are extracted from \cite{xxlbaryon}.
    \item \textbf{locuss\_highlx\_clusters.csv} - This contains information and properties for the clusters in the LoCuSS High-\lx{} sample, introduced in Section~\ref{subsubsec:LoCuSSsim}. A combination of overdensity radii, gas masses, and hydrostatic masses presented by \cite{locusshydro}, as well as positions, redshifts, \tx{}, and \lx{} measurements acquired via private communications.
\end{itemize}

We also provide a notebook that summarises the clusters in the samples; see Appendix~\ref{app:notebooks}. This notebook generates the final file in the this part of the companion repository, \textbf{combined\_sample\_duplicates.csv}, that details which of the clusters in each sample appear in another of the validation samples. This information is used in Section~\ref{subsec:samples} to state the number of unique clusters in our combined validation samples.

\subsection{Data Notes}
\label{app:datanotes}

A small, but important, part of this paper's companion repository, that includes notes made on the data used during analysis of the clusters in this work. This is relevant to all results presented in this paper. We include the following files:

\begin{itemize}[leftmargin=*, labelindent=2pt, itemsep=1ex]
\item \textbf{flare\_check\_notes.md} - We performed manual inspections of all the observations that \xga{} selected for the three samples used in this work, to check for evidence of residual flaring that could affect our results. This file contains the observation identifiers of the data that we decided to exclude, and in some cases provides extra context as to the reason.
\item \textbf{obs\_blacklist.csv} - This file contains the exact data we excluded from use, in a form that \xga{} recognises and can use as a `blacklist'.
\item \textbf{obs\_info.json} - Here we detail which \xmm{} observations are used in the analysis of which galaxy clusters. The file is human readable, and can be used to deduce which data are relevant to which clusters, but can also be read as a Python dictionary. The top-level keys are \xmm{} ObsIDs, and each is accompanied by a list of cluster names.
\end{itemize}

\subsection{Notebooks}
\label{app:notebooks}

\begin{itemize}[leftmargin=*, labelindent=2pt, itemsep=1ex]
\item \textbf{sample\_properties.ipynb} - This includes the information available in each dataset, the positions of the clusters on the sky, and whether there any common clusters between the samples. 
% It is also used to generate Figure~\ref{fig:samp_sum}.
\item \textbf{demonstration\_clusters.ipynb} - Here we select our two example clusters, \clustone{} and \clusttwo{}, and generate every figure included in Section~\ref{sec:methodology}. The information in Table~\ref{tab:exampclusters} is also shown here. 
\item \textbf{common.py} - This contains variable definitions that are common to multiple notebooks, so any change is reflected in the entire analysis once re-run. This includes the cosmology definitions and colour selections for our three validation samples. We also include comparison and plotting functions common to multiple notebooks, these are used to generate the one-to-one comparison figures in Section~\ref{sec:validation}.
\item \textbf{visualisations\_region\_adjustment} - 
This is a directory, containing several notebooks related to the visualisation of our clusters, and the modification of the relevant region files.
    \begin{itemize}[leftmargin=*, labelindent=2pt, itemsep=1ex]
        \item[+] \textbf{visualise\_analysis\_regions.ipynb} - In this notebook we create visualisations of the \xmm{} observations of the clusters in the three samples. We also overlay region information, and save the images as an output (see Appendix~\ref{app:outputs}).
        \item[+] \textbf{*\_region\_checks.ipynb} - This represents three notebooks (with the `*' replaced by `sdssrm-xcs', `xxl', and `locuss') that serve the same purpose for each of our three samples introduced in Section~\ref{subsec:samples}. Here we used the \xga{} region editor (see Table~\ref{tab:xgacomm}) to make manual adjustments to the regions used to remove contaminating sources from the analyses of each sample. These notebooks are relevant to Section~\ref{subsec:manualdetreg}.
    \end{itemize}

\item \textbf{temp\_lum\_comparisons} - We store notebooks relevant to the measurement of X-ray temperatures and luminosities for our three samples. These measurements are used for comparisons of \xga{} measurements to literature values, as seen in Section~\ref{subsubsec:txtests}. 
    \begin{itemize}[leftmargin=*, labelindent=2pt, itemsep=1ex]
        \item[+] \textbf{sdss\_comparisons.ipynb} - \xga{} measurements of temperature and luminosity, within both $R_{500}$ and $R_{2500}$ (where available), are performed here for the SDSSRM-XCS sample (Section~\ref{subsubsec:SDSSRMsim}).This notebook is used to create Figure~\ref{fig:xcstempcomp}. We also measure `core-excised' properties for SDSSRM-XCS.
        \item[+] \textbf{xxl\_comparisons.ipynb} - 
        \xga{} measurements of temperature and luminosity within a 300~kpc aperture for the XXL sample (Section~\ref{subsubsec:XXLsim}) are made in this notebook. We also create Figure~\ref{fig:xxltxcomp} by comparing to the original XXL temperature measurements \citep[][]{xxllt}.
        \item[+] \textbf{locuss\_comparisons.ipynb} - This notebook contains \xga{} measurements of temperature and luminosity for the LoCuSS High-\lx{} sample (Section~\ref{subsubsec:LoCuSSsim}). The comparisons made to literature values are shown in Figure~\ref{fig:locusstxcomp}.
    \end{itemize}

\item \textbf{gas\_mass\_comparisons} - A directory containing notebooks relevant to the comparison of literature gas mass measurements to \xga{} measurements, as seen in Section~\ref{subsec:compgasmasses}.
    \begin{itemize}[leftmargin=*, labelindent=2pt, itemsep=1ex]
        \item[+] \textbf{xxl\_gm\_comparisons.ipynb} - \xga{} measurements of gas mass for the XXL-100-GC sample within published $R_{500}$ (with radius error propagated) are performed here. Comparisons, as discussed in Section~\ref{subsec:compgasmasses}, are performed, and the notebook is used to generate Figure~\ref{fig:compxxlgm}.
        \item[+] \textbf{locuss\_gm\_comparisons.ipynb} - \xga{} measurements of gas mass for the LoCuSS High-\lx{} sample within published $R_500$ and $R_{2500}$ values are performed here. Comparisons, as discussed in Section~\ref{subsec:compgasmasses}, are performed, and the notebook is used to generate Figure~\ref{fig:complocussgm}.
    \end{itemize}

\item \textbf{hydro\_masses}
    Finally, this directory contains notebooks relevant to the measurement of hydrostatic masses for two of the cluster samples used in this work.
    \begin{itemize}[leftmargin=*, labelindent=2pt, itemsep=1ex]
        \item[+] \textbf{locuss\_hym\_comparisons.ipynb} - \xga{} measurements of hydrostatic mass for LoCuSS High-\lx{} galaxy clusters are performed here, within both $R_{500}$ and $R_{2500}$ radii. In Section~\ref{subsec:mhydrocomp} they are compared to the literature values measured using \xmm{}. This notebook is used to generate Figure~\ref{fig:complocusshym}.
        \item[+] \textbf{sdssrm-xcs\_new\_masses.ipynb} - In this notebook we measure new hydrostatic masses for the SDSSRM-XCS sample of galaxy clusters, as presented in Section~\ref{sec:results}.
    \end{itemize}

\end{itemize}

\subsection{Outputs}
\label{app:outputs}

In this final part of the companion repository, we store the main outputs of this work. 

\begin{itemize}[leftmargin=*, labelindent=2pt, itemsep=1ex]
\item \textbf{cluster\_visualisations} - This directory contains combined \xmm{} images of the clusters in the three samples, with one sub-directory for each sample. The file name used for each image is the name of the galaxy cluster used in this work, a mask has been applied, and contaminating source regions are overlaid. 
\item \textbf{custom\_regions} - This contains region files which were modified during the course of Section~\ref{subsec:manualdetreg}. They are in sky (RA-Dec) coordinates, and in the FK5 coordinate frame.
\item \textbf{figures} - Figures created for this paper are stored in this directory.
\item \textbf{results} - All sets of results measured during the course of this work are stored in here, including \xga{} measured properties for the SDSSRM-XCS, XXL-100-GC, and LoCuSS High-\lx{} samples. Tables~\ref{tab:sdssrmmeasurements}, \ref{tab:xxlmeasurements}, \ref{tab:locussmeasurementsA}, and \ref{tab:locussmeasurementsB} contain a subset of the results, and define the columns.
\end{itemize}

\section{Tables of properties}

\subsection{SDSSRM-XCS \xga{} properties}

\begin{table*}
%comment: This won't contain all the results, I just don't think it'll be a good use of space. And there are quite a few tables that I want to include DT
\begin{center}
\caption[]{{Properties of the SDSSRM-XCS galaxy cluster sample.  All uncertainties calculated from 68\% confidence limits, equivalent to $1\sigma$.  For brevity, the first entry of the Table is given, with the full Table available from \href{https://github.com/DavidT3/XCS-Mass-Paper-I-Analysis/tree/master/outputs/results}{https://github.com/DavidT3/XCS-Mass-Paper-I-Analysis/tree/master/outputs/results}.\\  (1) redMaPPer ID of the cluster; (2) Right ascension as defined by redMaPPer; (3) Declination taken from redMaPPer; (4) redshift taken from redMaPPer; (5) $R_{500}$ taken from \cite{xcsgiles}; (6) $R_{2500}$ taken from \cite{xcsgiles}; (7) X-ray Temperature determined within $R_{2500}$; (8) X-ray temperature determined within $R_{500}$; (9) Gas mass determined within $R_{2500}$; (10) Gas mass determined within $R_{500}$; (11) Mass determined from the hydrostatic mass analysis described throughout Section~\ref{sec:methodology}, within $R_{2500}$; (12) Mass determined from the hydrostatic mass analysis described throughout Section~\ref{sec:methodology}, within $R_{500}$.
\label{tab:sdssrmmeasurements}}}
\vspace{1mm}
\begin{tabular}{cccccccccccc}
\hline
\hline
(1) & (2) & (3) & (4) & (5) & (6) & (7) & (8) & (9) & (10) & (11) & (12) \\
RM ID & RA & Dec & z & $R_{500}$ & $R_{2500}$ & $T_{\rm{X}}^{2500}$ & $T_{\rm{X}}^{500}$ & $M^{\rm{gas}}_{2500}$ & $M^{\rm{gas}}_{500}$ & $M^{\rm{hy}}_{2500}$ & $M^{\rm{hy}}_{500}$\\
 & (deg) & (deg) & & (kpc) & (kpc) & (keV) & (keV) & 10$^{14}$ (M$_{\odot}$) & 10$^{14}$ (M$_{\odot}$) & 10$^{14}$ (M$_{\odot}$) & 10$^{14}$ (M$_{\odot}$) \\
\hline
\hline
2 & 250.08 & 46.71 & 0.23 & 1503 & 644 & 9.73$^{+0.20}_{-0.20}$ & 10.40$^{+0.24}_{-0.22}$ & 0.59$^{+0.00}_{-0.00}$ & 1.68$^{+0.00}_{-0.00}$ & 4.48$^{+0.26}_{-0.26}$ & 10.36$^{+1.31}_{-1.34}$ \\
\hline
\end{tabular}
\end{center}
\end{table*}

\subsection{XXL-100-GC \xga{} Reanalysis Results}
\label{app:xxlresults}

\begin{table*}
\begin{center}
\caption[]{{Properties of the XXL-100-GC galaxy cluster sample.  All uncertainties calculated from 68\% confidence limits, equivalent to $1\sigma$.  For brevity, the first entry of the Table is given, with the full Table available from \href{https://github.com/DavidT3/XCS-Mass-Paper-I-Analysis/tree/master/outputs/results}{https://github.com/DavidT3/XCS-Mass-Paper-I-Analysis/tree/master/outputs/results}.\\ (1) XLSSC ID as given in \cite{gc100}; (2) Right ascension as given in \cite{gc100}; (3) Declination as given in \cite{gc100}; (4) redshift as given in \cite{gc100}; (5) $R_{500}$ as given in \cite{xxlbaryon}; (6) \xga{} measured X-ray temperature within 300~kpc; (7) \xga{} measured gas mass within $R_{500}$.}
\label{tab:xxlmeasurements}}
\vspace{1mm}
\begin{tabular}{ccccccccccc}
\hline
\hline
(1) & (2) & (3) & (4) & (5) & (6) & (7) \\
XLSSC & RA & Dec & z & $R_{500}$ & $T_{\rm{X}}^{\rm 300kpc}$ & $M^{\rm{gas}}_{500}$ \\
 & (deg) & (deg) & & (kpc) & (keV) & 10$^{14}$ (M$_{\odot}$) \\
\hline
\hline
1 & 36.24 & -3.82 & 0.614 & 777 & 3.41$^{+0.47}_{-0.38}$ & 0.31$^{+0.07}_{-0.07}$ \\
\hline
\end{tabular}
\end{center}
\end{table*}

\subsection{L\lowercase{o}C\lowercase{u}SS High-$L_{\rm{X}}$ \xga{} Reanalysis Results}
\label{app:locussresults}

\begin{table*}
\begin{center}
\caption[]{{LoCuSS High-$L_{\rm{X}}$ (a) galaxy cluster properties.  
All uncertainties calculated from 68\% confidence limits, equivalent to $1\sigma$. For brevity, the first entry of the Table is given, with the full Table available from \href{https://github.com/DavidT3/XCS-Mass-Paper-I-Analysis/tree/master/outputs/results}{https://github.com/DavidT3/XCS-Mass-Paper-I-Analysis/tree/master/outputs/results}.\\  (1) Cluster name; (2) Right ascension; (3) Declination; (4) redshift; (5) $R_{2500}$ overdensity radius; (6) $R_{500}$ overdensity radius; (7) \xga{} measured gas mass within $R_{2500}$; (8) \xga{} measured gas mass within $R_{500}$; (9) \xga{} measured hydrostatic mass within $R_{2500}$; (10) \xga{} measured hydrostatic mass within $R_{500}$. Values for columns (2) - (6) are taken from \cite{locusshydro}.}
\label{tab:locussmeasurementsA}}
\vspace{1mm}
\begin{tabular}{cccccccccc}
\hline
\hline
(1) & (2) & (3) & (4) & (5) & (6) & (7) & (8) & (9) & (10) \\
Name & RA & Dec & z & $R_{2500}$ & $R_{500}$ & $M^{\rm{gas}}_{2500}$ & $M^{\rm{gas}}_{500}$ & $M^{\rm{hy}}_{2500}$ & $M^{\rm{hy}}_{500}$ \\
 & (deg) & (deg) & & (kpc) & (kpc) & 10$^{14}$ (M$_{\odot}$) & 10$^{14}$ (M$_{\odot}$) & 10$^{14}$ (M$_{\odot}$) & 10$^{14}$ (M$_{\odot}$) \\
\hline
\hline
Abell 0068 & 9.2785 & 9.1566 & 0.255 & 580 & 1400 & 0.32$^{+0.00}_{-0.00}$ & 0.83$^{+0.00}_{-0.00}$ & 3.79$^{+0.18}_{-0.18}$ & 9.68$^{+0.99}_{-0.97}$ \\
\hline
\end{tabular}
\end{center}
\end{table*}

\begin{table*}
\begin{center}
\caption[]{{Properties of the LoCuSS High-$L_{\rm{X}}$ (b) galaxy cluster sample.  All uncertainties calculated from 68\% confidence limits, equivalent to $1\sigma$. For brevity, the first entry of the Table is given, with the full Table available from \href{https://github.com/DavidT3/XCS-Mass-Paper-I-Analysis/tree/master/outputs/results}{https://github.com/DavidT3/XCS-Mass-Paper-I-Analysis/tree/master/outputs/results}.\\ (1) Cluster name; (2) Right ascension; (3) Declination; (4) redshift; (5) $R_{500}$ overdensity radius determined from a weak lensing analysis \citep[see][values were provided by G. Smith via priv. comm.]{locusswl}; (6) \xga{} measured core-excluded X-ray temperature within [0.15-1]$R_{\rm WL,500}$.}
\label{tab:locussmeasurementsB}}
\vspace{1mm}
\begin{tabular}{cccccc}
\hline
\hline
(1) & (2) & (3) & (4) & (5) & (6) \\
Name & RA & Dec & z & $R_{{\rm WL},500}$ & $T_{\rm{X}}^{\rm 500ce}$ \\
 & (deg) & (deg) & & (kpc) & (keV) \\
\hline
\hline
Abell 0068 & 9.2785 & 9.1566 & 0.255 & 1231 & 7.33$^{+0.12}_{-0.12}$ \\
\hline
\end{tabular}
\end{center}
\end{table*}

\section{Relevant \xga{} commands and classes}
\label{app:xgacommclass}

\begin{table*}
\begin{center}
\caption[]{The \xga{} product classes used in this work (with a link to their API documentation), where they can be found in the hierarchy of \xga{} \verxga{}, a brief summary of their purpose and capabilities, and the most relevant section of this paper that they were used in. \label{tab:xgaclass}}
\vspace{1mm}
\begin{tabular}{|l|p{0.14\textwidth}|p{0.5\textwidth}|c|} 
\hline
\hline
Link to \xga{} Documentation & \xga{} Hierarchy & Summary & Section\\
\hline
\hline

\href{https://xga.readthedocs.io/en/v0.4.3/xga.products.html\#xga.products.phot.Image}{\texttt{Image}}
&
\texttt{products.phot}
&
Provides an \xga{} interface to FITS images, with easy access to the data and headers (read into memory only when needed), and convenient methods for coordinate conversion and creating visualisations.
&
\S~\ref{subsec:xgaphot}
\\\hline

\href{https://xga.readthedocs.io/en/v0.4.3/xga.products.html\#xga.products.phot.ExpMap}{\texttt{ExpMap}}
&
\texttt{products.phot}
&
Identical in function to \texttt{Image}, but provides an interface for exposure maps.
&
\S~\ref{subsec:xgaphot}
\\\hline

\href{https://xga.readthedocs.io/en/v0.4.3/xga.products.html\#xga.products.phot.RateMap}{\texttt{RateMap}}
&
\texttt{products.phot}
&
Provides an interface to count-rate maps, constructed from \texttt{Image} and \texttt{ExpMap} instances. Has all \texttt{Image} functionality, as well as extra methods to calculate signal-to-noise within a specified region.
&
\S~\ref{subsec:xgaphot}
\\\hline

\texttt{\href{https://xga.readthedocs.io/en/v0.4.3/xga.products.html\#xga.products.spec.Spectrum}{Spectrum}}
&
\tt{products.spec}
&
Provides an \xga{} interface to FITS spectra, with access to data and headers, as well as the ancillary files (background spectra, ARF, and RMF). Can store model fit information, and create visualisations of the spectra and effective area curve.
&
\S~\ref{subsec:xgaspecs}
\\\hline

\texttt{\href{https://xga.readthedocs.io/en/v0.4.3/xga.products.html\#xga.products.spec.AnnularSpectra}{AnnularSpectra}}
&
\tt{products.spec}
&
Constructed from a set of \texttt{Spectrum} objects which have been generated in concentric annuli, this product provides similar access to data and model fits. Visualisation methods are also included, to view spectra from multiple annuli at once or view all spectra for the same annulus. 
&
\S~\ref{subsec:xgaspecs}
\\\hline

\texttt{\href{https://xga.readthedocs.io/en/v0.4.3/xga.products.html\#xga.products.profile.ProjectedGasTemperature1D}{ProjectedGasTemperature1D}}
&
\texttt{products.profile}
&
Contains projected temperature profile information (i.e. the temperatures measured by fitting the \texttt{AnnularSpectra} components with \xspec{}). All \xga{} profiles have the ability to fit \xga{} models to their data, produce visualisations, and provide easy access to their data.
&
\S~\ref{subsec:xgaspecs}
\\\hline

\texttt{\href{https://xga.readthedocs.io/en/v0.4.3/xga.products.html\#xga.products.profile.APECNormalisation1D}{APECNormalisation1D}}
&
\texttt{products.profile}
&
Similar to the \texttt{ProjectedGasTemperature1D} profile, but containing APEC normalisations from the fitting of an \texttt{AnnularSpectra}.
&
\S~\ref{subsec:xgaspecs}
\\\hline

\texttt{\href{https://xga.readthedocs.io/en/v0.4.3/xga.products.html\#xga.products.profile.EmissionMeasure1D}{EmissionMeasure1D}}
&
\texttt{products.profile}
&
An \xga{} profile class for emission measure. They can be produced from \texttt{APECNormalisation1D} profiles. 
&
\S~\ref{subsec:xgaspecs}
\\\hline

\texttt{\href{https://xga.readthedocs.io/en/v0.4.3/xga.products.html\#xga.products.profile.GasDensity3D}{GasDensity3D}} 
&
\tt{products.profile}
&
\xga{} profile class for the 3D hot gas density (either number or mass density). Can calculate a gas mass within a specified radius, assuming spherical symmetry.
&
\S~\ref{subsec:densityProf}
\\\hline

\texttt{\href{https://xga.readthedocs.io/en/v0.4.3/xga.products.html\#xga.products.profile.GasTemperature3D}{GasTemperature3D}} 
&
\tt{products.profile}
&
\xga{} profile class for deprojected (i.e. three-dimensional) hot gas temperatures.
&
\S~\ref{subsec:gastempprof}
\\\hline

\texttt{\href{https://xga.readthedocs.io/en/v0.4.3/xga.products.html\#xga.products.profile.HydrostaticMass}{HydrostaticMass}} 
&
{\tt products.profile}
&
Represents a hydrostatic mass profile calculated from a \texttt{GasTemperature3D} and \texttt{GasDensity3D} profile. Contains extra methods to calculate a hydrostatic mass at a particular radius, as well as visualising the posterior distribution.
&
\S~\ref{subsec:hydromassprof}
\\\hline

\texttt{\href{https://xga.readthedocs.io/en/v0.4.3/xga.models.html\#xga.models.sb.BetaProfile1D}{BetaProfile1D}} 
&
\tt{models.sb}
&
\xga{} model class for the simplest model of a galaxy cluster X-ray surface brightness profile. These model classes can create basic visualisations of the current shape of the model, store parameter values and distributions, and perform mathematical tasks like differentiation, spherical integration, and Abel transforms.
&
\S~\ref{subsec:SBandEm}
\\\hline

\texttt{\href{https://xga.readthedocs.io/en/v0.4.3/xga.models.html\#xga.models.density.KingProfile1D}{KingProfile1D}} 
&
\tt{models.density}
&
\xga{} model class for the simplest model of a galaxy cluster 3D density profile, deriving from the \texttt{BetaProfile1D}. 
&
\S~\ref{subsec:densityProf}
\\\hline

\texttt{\href{https://xga.readthedocs.io/en/v0.4.3/xga.models.html\#xga.models.temperature.SimpleVikhlininTemperature1D}{SimpleVikhlininTemperature1D}} 
&
\tt{models.temperature}
&
\xga{} model class for a simplified version of the Vikhlinin model for the 3D temperature profile of the ICM.
&
\S~\ref{subsec:gastempprof}
\\\hline

\end{tabular}

\end{center}
\end{table*}

% \texttt{\href{https://xga.readthedocs.io/en/v0.4.3/xga.products.html\#xga.products.profile.GasMass1D}{GasMass1D}}
% &
% \tt{products.profile}
% &
% Cumulative gas mass profile product, which can be generated from a \texttt{GasMass1D}
% &
% \S~\ref{subsec:densityProf}
% \\\hline

% Settings Used column was removed - maybe I'll re-introduce it but I'm not sure 

\begin{table*}
\begin{center}
\caption[]{\xga{} methods and functions used in this work (with a link to their API documentation), where they can be found in the hierarchy of \xga{} \verxga{}, a brief summary of their purpose and capabilities, and the most relevant section of this paper that they were used in. This is not an exhaustive list of every function called in the course o these analyses, but are the most salient to understanding what we did with \xga{} to achieve our results. \label{tab:xgacomm}}
\vspace{1mm}
\begin{tabular}{|l|p{0.18\textwidth}|p{0.48\textwidth}|c|}\\ 
\hline
\hline
Link to \xga{} Documentation & \xga{} Hierarchy & Summary & Section \\
\hline
\hline

\href{https://xga.readthedocs.io/en/v0.4.3/xga.sourcetools.html#xga.sourcetools.match.simple_xmm_match}{\texttt{simple\_xmm\_match}}
&
\tt{sourcetools.match}
&
Returns ObsIDs within a specified distance from the input $\alpha$ and $\delta$ values.
&
\S~\ref{subsec:xcsdata}\\\hline

\texttt{\href{https://xga.readthedocs.io/en/v0.4.3/xga.products.html\#xga.products.phot.Image.edit_regions}{edit\_regions}} 
&
\texttt{products.phot.Image}
& 
Method of \texttt{Image} to show an interactive image to add and modify source regions.
&
\S~\ref{subsec:manualdetreg}\\\hline

\href{https://xga.readthedocs.io/en/v0.4.3/xga.products.html\#xga.products.phot.Image.view}{\texttt{view}}
&
\texttt{products.phot.Image}
&
Make visualisations of \texttt{Image}, \texttt{ExpMap}, and \texttt{RateMap}, built into the classes. Can be configured to add overlays and masks.
&
\S~\ref{subsec:xgaphot}\\\hline

\texttt{\href{https://xga.readthedocs.io/en/v0.4.3/xga.imagetools.html\#xga.imagetools.profile.radial_brightness}{radial\_brightness}}
&
\tt{imagetools.profile}
&
Constructs radial surface brightness profiles from concentric circular annuli.
&
\S~\ref{subsec:SBandEm}\\\hline

\href{https://xga.readthedocs.io/en/v0.4.3/xga.products.html\#xga.products.base.BaseProfile1D.fit}{\texttt{fit}}
&
\texttt{products.base. BaseProfile}
&
A method of all \xga{} profile classes, which can fit models to their data and store the resulting information within the profile.
&
App.~\ref{app:radmodfit}\\\hline

\texttt{\href{https://xga.readthedocs.io/en/v0.4.3/xga.sources.html\#xga.sources.extended.GalaxyCluster.norm_conv_factor}{norm\_conv\_factor}} 
&
{\tt sources.extended. GalaxyCluster}
&
Calculates count-rate to APEC normalisation conversion factors for use on surface brightness profiles (both individual instrument and combined).
&
\S~\ref{subsec:densityProf}\\\hline

\texttt{\href{https://xga.readthedocs.io/en/v0.4.3/xga.sourcetools.html\#xga.sourcetools.deproj.shell_ann_vol_intersect}{shell\_ann\_vol\_intersect}} 
&
\tt{sourcetools.deproj}
&
Calculates the volume intersection matrix of a set of circular annuli and a set of spherical shells
&
\S~\ref{subsec:gastempprof}\\\hline

\texttt{\href{https://xga.readthedocs.io/en/v0.4.3/xga.xspec.fit.html\#xga.xspec.fit.general.single_temp_apec}{single\_temp\_apec}}
&
\tt{xspec.fit.general}
&
A function which fits emission models to spectra (either in a circular aperture or a single annulus) in order to measure `global properties'. Information on inner and outer radii is taken as an argument, to retrieve the correct spectra or generate them.
&
\S~\ref{subsec:xgaspecs}\\\hline

\texttt{\href{https://xga.readthedocs.io/en/v0.4.3/xga.xspec.fit.html\#xga.xspec.fit.profile.single_temp_apec_profile}{single\_temp\_apec\_profile}}
&
\tt{xspec.fit.profile}
&
A function which fits emission models to a set of annular spectra (taking information about the fit configuration, as well as the size and number of the annuli so that it can generate them if they do not exist). This function specifically fits absorbed single-temperature APEC models.
&
\S~\ref{subsec:xgaspecs}\\\hline

\texttt{\href{https://xga.readthedocs.io/en/v0.4.3/xga.sourcetools.html\#xga.sourcetools.mass.inv_abel_dens_onion_temp}{inv\_abel\_dens\_onion\_temp}} 
&
\tt{sourcetools.mass}
&
Calls other \xga{} functions (see the next two entries) to generate 3D density and temperature profiles, then combines them into hydrostatic mass profiles.
&
\S~\ref{subsec:hydromassprof}\\\hline

\texttt{\href{https://xga.readthedocs.io/en/v0.4.3/xga.sourcetools.html\#xga.sourcetools.temperature.onion_deproj_temp_prof}{onion\_deproj\_temp\_prof}} 
&
\tt{sourcetools.temperature}
&
This function infers the 3D temperature profile from a projected temperature profile via the onion peeling method. If no projected temperature profiles exist they are generated by calling \texttt{\href{https://xga.readthedocs.io/en/v0.4.3/xga.sourcetools.html\#xga.sourcetools.temperature.min_cnt_proj_temp_prof}{min\_cnt\_proj\_temp\_prof}}.
&
\S~\ref{subsec:gastempprof}\\\hline

\texttt{\href{https://xga.readthedocs.io/en/v0.4.3/xga.sourcetools.html\#xga.sourcetools.temperature.min_cnt_proj_temp_prof}{min\_cnt\_proj\_temp\_prof}} 
&
\tt{sourcetools.temperature}
&
Uses a user-specified minimum number of counts and 
minimum annulus width to define annular regions and generate a projected temperature profile. This is achieved by calling \texttt{\href{https://xga.readthedocs.io/en/v0.4.3/xga.xspec.fit.html\#xga.xspec.fit.profile.single_temp_apec_profile}{single\_temp\_apec\_profile}}, which will both ensure that the spectra have been generated, and fit them with emission models.
&
\S~\ref{subsec:gastempprof}\\\hline

\texttt{\href{https://xga.readthedocs.io/en/v0.4.3/xga.sourcetools.html\#xga.sourcetools.density.inv_abel_fitted_model}{inv\_abel\_fitted\_model}} 
&
\tt{sourcetools.density}
&
Generates 3D gas density profiles by the surface-brightness profile method, converting them to 3D emissivity via an inverse abel transform, and from there to density by a count-rate to APEC normalisation factor.
&
\S~\ref{subsec:densityProf}\\\hline

\end{tabular}

\end{center}
\end{table*}

%%%%%%%%%%%%%%%%%%%%%%%%%%%%%%%%%%%%%%%%%%%%%%%%%%

% Don't change these lines
\bsp	% typesetting comment
\label{lastpage}
\end{document}